\documentclass[pdftex]{pasj01}
\Received{$\langle$reception date$\rangle$}
\Accepted{$\langle$acception date$\rangle$}
\Published{$\langle$publication date$\rangle$}
\usepackage{lscape,dcolumn} % 2017.4.11 HN
\usepackage{aas_macros} % 2017.6.20
\usepackage{graphicx} % 2017.9.26
\begin{document}

\title{In-orbit performance of the soft X-ray imaging system aboard Hitomi (ASTRO-H)
\thanks{The corresponding authors are
Hiroshi \textsc{Nakajima},
Yoshitomo \textsc{Maeda},
Hiroyuki \textsc{Uchida},
and
Takaaki \textsc{Tanaka}
}
}
\author{Hiroshi \textsc{Nakajima}\altaffilmark{1,2},
Yoshitomo \textsc{Maeda}\altaffilmark{3},
Hiroyuki \textsc{Uchida}\altaffilmark{4},
Takaaki \textsc{Tanaka}\altaffilmark{4},
Hiroshi \textsc{Tsunemi}\altaffilmark{1,2},
Kiyoshi \textsc{Hayashida}\altaffilmark{1,2},
Takeshi G. \textsc{Tsuru}\altaffilmark{4},
Tadayasu \textsc{Dotani}\altaffilmark{3},
%Naohisa \textsc{Anabuki}\altaffilmark{1,2},
Ryo \textsc{Nagino}\altaffilmark{1,2,20},
Shota \textsc{Inoue}\altaffilmark{1},
Masanobu \textsc{Ozaki}\altaffilmark{3},
Hiroshi \textsc{Tomida}\altaffilmark{3},
Chikara \textsc{Natsukari}\altaffilmark{3},
Shutaro \textsc{Ueda}\altaffilmark{3},
Koji \textsc{Mori}\altaffilmark{5},
Makoto \textsc{Yamauchi}\altaffilmark{5},
Isamu \textsc{Hatsukade}\altaffilmark{5},
Yusuke \textsc{Nishioka}\altaffilmark{5},
Miho \textsc{Sakata}\altaffilmark{5},
Tatsuhiko \textsc{Beppu}\altaffilmark{5},
Daigo \textsc{Honda}\altaffilmark{5},
Masayoshi \textsc{Nobukawa}\altaffilmark{6},
Junko S. \textsc{Hiraga}\altaffilmark{7},
Takayoshi \textsc{Kohmura}\altaffilmark{8},
Hiroshi \textsc{Murakami}\altaffilmark{9},
Kumiko K. \textsc{Nobukawa}\altaffilmark{10},
Aya \textsc{Bamba}\altaffilmark{11},
John P. \textsc{Doty}\altaffilmark{12},
Ryo \textsc{Iizuka}\altaffilmark{3},
Toshiki \textsc{Sato}\altaffilmark{3},
Sho \textsc{Kurashima}\altaffilmark{3},
Nozomi \textsc{Nakaniwa}\altaffilmark{3},
Ryota \textsc{Asai}\altaffilmark{3},
Manadu \textsc{Ishida}\altaffilmark{3},
Hideyuki \textsc{Mori}\altaffilmark{13},
Yang \textsc{Soong}\altaffilmark{13},
Takashi \textsc{Okajima}\altaffilmark{13},
Peter \textsc{Serlemitsos}\altaffilmark{13},
Yuzuru \textsc{Tawara}\altaffilmark{14},
Ikuyuki \textsc{Mitsuishi}\altaffilmark{14},
Kazunori \textsc{Ishibashi}\altaffilmark{14},
Keisuke \textsc{Tamura}\altaffilmark{14},
Takayuki \textsc{Hayashi}\altaffilmark{14},
Akihiro \textsc{Furuzawa}\altaffilmark{15},
Satoshi \textsc{Sugita}\altaffilmark{16},
Takuya \textsc{Miyazawa}\altaffilmark{17},
Hisamitsu \textsc{Awaki}\altaffilmark{18},
Eric D. \textsc{Miller}\altaffilmark{19},
Hiroya \textsc{Yamaguchi}\altaffilmark{13}
}
\altaffiltext{1}{Department of Earth and Space Science, Osaka University, 1-1 Machikaneyama-cho, Toyonaka, Osaka 560-0043}
\altaffiltext{2}{Project Research Center for Fundamental Sciences, Osaka University, 1-1 Machikaneyama-cho, Toyonaka, Osaka 560-0043}
\altaffiltext{3}{Japan Aerospace Exploration Agency, Institute of Space and Astronautical Science, 3-1-1 Yoshino-dai, Chuo-ku, Sagamihara, Kanagawa 252-5210}
\altaffiltext{4}{Department of Physics, Kyoto University, Kitashirakawa Oiwake-cho, Sakyo,
Kyoto, Kyoto 606-8502}
\altaffiltext{5}{Faculty of Engineering, University of Miyazaki, 1-1 Gakuen Kibanadai Nishi,
Miyazaki 889-2192}
\altaffiltext{6}{Faculty of Education, Nara University of Education, Takabatake-cho, Nara, Nara 630-8528}
\altaffiltext{7}{Department of Physics, Kwansei Gakuin University, 2-2 Gakuen, Sanda,
Hyogo 669-1337}
\altaffiltext{8}{Department of Physics, Tokyo University of Science, 2641 Yamazaki, Noda,
Chiba 270-8510}
\altaffiltext{9}{Faculty of Liberal Arts, Tohoku Gakuin University, 2-1-1 Tenjinzawa, Izumi-ku, Sendai, Miyagi 981-3193}
\altaffiltext{10}{Department of Physics, Nara Women's University, Kitauoyanishi-machi, Nara, Nara 630-8506}
\altaffiltext{11}{Department of Physics, University of Tokyo, 7-3-1 Hongo, Bunkyo, Tokyo 113-0033}
\altaffiltext{12}{Noqsi Aerospace Ltd, 2822 S Nova Road, Pine, CO 80470, USA}
\altaffiltext{13}{NASA's Goddard Space Flight Center, 8800 Greenbelt Road, Greenbelt, MD 20771, USA}
\altaffiltext{14}{Department of Physics, Nagoya University, Furo-cho, Chikusa-ku, Nagoya, Aichi 464-8602}
\altaffiltext{15}{Fujita Health University, Toyoake, Aichi 470-1192}
\altaffiltext{16}{Department of Physics, Tokyo Institute of Technology, 2-12-1 Ookayama, Meguro-ku, Tokyo 152-8550}
\altaffiltext{17}{Okinawa Institute of Science and Technology Graduate University, 1919-1 Tancha, Onna-son Okinawa, 904-0495}
\altaffiltext{18}{Department of Physics, Ehime University, Bunkyo-cho, Matsuyama, Ehime 790-8577}
\altaffiltext{19}{Kavli Institute for Astrophysics and Space Research, Massachusetts Institute of Technology, 77 Massachusetts Avenue, Cambridge, MA 02139, USA}
\altaffiltext{20}{NEC Aerospace Systems, Ltd., 5-22-5 Sumiyoshi-cho, Fuchu, Tokyo 183-8502}
\email{nakajima@ess.sci.osaka-u.ac.jp}


\maketitle

\begin{abstract}
We describe the in-orbit performance of the soft X-ray imaging system
consisting of the Soft X-ray Telescope and the Soft X-ray Imager aboard Hitomi.
Verification and calibration of imaging and spectroscopic
performance are carried out making the best use of the limited data
of less than three weeks.
%\textcolor{blue}{{\bf Comprehensive functionality checks
%such as the cooling capability and the charge injection are followed by
%the in-orbit calibration of imaging and spectroscopic performance indices.}}
Basic performance including a large field of view of
\timeform{38'}~$\times$~\timeform{38'} is verified
with the first light image of the Perseus cluster of galaxies.
%A large field of view of \timeform{38'}~$\times$~\timeform{38'} is verified
%with the first light image of the Perseus cluster of galaxies.
Amongst the small number of
observed targets, the on-minus-off pulse image for the out-of-time
events of the Crab pulsar enables us to measure a half power diameter
of the telescope as $\sim$~\timeform{1.3'}.
The average energy resolution measured with the onboard calibration source events
at 5.89~keV is 179~$\pm$~3~eV in full width at half maximum.
Light leak and cross talk issues affected the effective exposure time
and the effective area, respectively, because all the observations were
performed before optimizing an observation schedule and
parameters for the dark level calculation.
Screening the data affected by
these two issues, we measure the background level to be
5.6~$\times$~10$^{-6}$~counts~s$^{-1}$~arcmin$^{-2}$~cm$^{-2}$ in the energy
band of 5--12~keV, which is seven times lower than that of the Suzaku XIS-BI.
\end{abstract}

%\linenumbers
\section{Introduction}\label{sec:intro}

%Since the first true X-ray imaging by sounding rocket experiments
%with Wolter-I type imaging telescope payload \citep{1979ApJ...227..285R},
Extra-solar X-ray astronomy has advanced with
advent and utilization of the state-of-the-art instruments.
Key features of the instruments are imaging and spectroscopic
capabilities and wide-band coverage. The X-ray imaging capability
jumped up when the first Wolter-I type telescope was launched
by the sounding rocket \citep{1979ApJ...227..285R}.
Further improvement was achieved with
%a series of space missions with the state-of-the-art instruments.
Einstein \citep{1979ApJ...230..540G} and ROSAT \citep{1982AdSpR...2..241T}.
Energy resolution increased with the adoption of an X-ray CCD,
which was first realized by ASCA \citep{1994PASJ...46L..37T}
simultaneously extending the energy band up to 10~keV.
This achievement was followed by
%, and BeppoSAX \citep{1997A&AS..122..299B}
%realized both imaging and spectroscopy by the Wolter-I type
%telescopes and the gaseous detectors.
Chandra \citep{2002PASP..114....1W}, XMM-Newton \citep{2000AAS...196.3401J},
Swift \citep{2005SSRv..120..165B} and Suzaku \citep{2007PASJ...59S...1M}.
Wide band coverage was at first realized combining soft X-ray telescopes
and hard X-ray detectors like those aboard BeppoSAX \citep{1997A&AS..122..299B},
Suzaku and ASTROSAT \citep{2014SPIE.9144E..1SS}.
%enabled the high-resolution imaging and/or
%high-throughput spectroscopy in the soft X-ray band with CCDs.
%The recent launch of
NuSTAR \citep{2013ApJ...770..103H}
%and ASTROSAT \citep{2014SPIE.9144E..1SS}
broadened the energy range of X-ray telescope up to $\sim$~80~keV,
which was achieved using mirrors with
depth-graded multilayer coating and CdZnTe detectors.
In this context, ASTRO-H \citep{2016SPIE.9905E..0UT}
increased the energy resolution drastically with the wide band coverage.
%or down to EUV for the sensitive imaging.

%Soft X-ray Imaging system aboard Hitomi
ASTRO-H, the Japan-led X-ray observatory was
launched on February 17th, 2016 from Tanegashima Space Center.
It was successfully put into a planned low earth orbit (LEO)
and named Hitomi that is a Japanese word
for the pupil of the eye.
It is characterized by the wide-band imaging spectroscopy
from soft X-rays (0.3~keV) to soft gamma-rays (600~keV) and
high-resolution non-dispersive spectroscopy in soft X-rays.
Soft X-ray imaging system, one of the
four mission instruments, consists of the Soft X-ray Telescope
(SXT-I) (Serlemitsos et al. in preparation)
and the Soft X-ray Imager (SXI) \citep{Tanaka17}.

%paper structure
The detailed specification and the ground calibration of the SXT-I and the SXI
are described in other papers
(\cite{Iizuka17, Tanaka17, 2016SPIE.9905E..5DH, 2016NIMPA.831..415I, 2016OExpr..2425548K, 2016SPIE.9905E..3YK, 2016SPIE.9905E..3ZM, 2016JATIS...2d4001S, 2016SPIE.9905E..3XS, 2014SPIE.9144E..58I, 2014SPIE.9144E..59S, 2013NIMPA.731..160M}).
This paper focuses on the in-orbit
performance of the soft X-ray imaging system achieved
by fully utilizing the data obtained in the unexpectedly short mission life.
After describing the soft X-ray imaging system and initial operations
in section~\ref{sec:description} and \ref{sec:function}, respectively,
we summarize our observations in section~\ref{sec:observation} and
note two issues that affect the imaging and spectroscopic performances
in section~\ref{sec:issues}. The primary calibration items are highlighted
in section~\ref{sec:cal}, followed by summary in section~\ref{sec:summary}.
Errors quoted in this paper are in 90\% confidence level.

\section{The Soft X-ray Imaging System}\label{sec:description}

%mirror
The basic design of the
SXT-I (Serlemitsos et al. in preparation)
is the same as that of Suzaku XRT
\citep{2007PASJ...59S...9S}.
Thin Al foils with Au coated surfaces
are nested in a confocal way
to form a conically-approximated Wolter-I type optics.
This type of telescope provides relatively high effective area
per unit mass compared with the accurate Wolter-I type telescope
aboard Chandra \citep{2012OptEn..51a1013W}. Improvement of
the uniformity of the housing quality and alignment structure,
and the consistent
quality of the flight reflectors lead to remarkably better
image quality than that of Suzaku XRT.
The on-axis angular resolution of the SXT-I in the 1.5--17.5~keV
band was measured to be \timeform{1.3'}--\timeform{1.5'}
in a half power diameter (HPD) by the on-ground calibration
\citep{Iizuka17}.

%detector (SXI)
The SXI adopts P-channel back-illumination
type CCDs manufactured by Hamamatsu Photonics K. K. \citep{Tanaka17}.
The thin surface dead layer allows high quantum efficiency in soft X-rays
and high tolerance against impacts of micrometeoroids.
The depletion layer of 200~$\mu$m \citep{2008SPIE.7011E..0QT} is
substantially thicker than that of Suzaku XIS-BI of
42~$\mu$m \citep{2007PASJ...59S..23K},
which enhances quantum efficiency for X-rays
above 6~keV. Four CCDs (hereafter CCD1, CCD2, CCD3 and CCD4, respectively)
are placed in a 2~$\times$~2 array at
the focal plane of the SXT-I and form an imaging area with a size
of $\sim$~60~mm~$\times$~60~mm.

%SXI continuued
The SXI consists of four components: the sensor (SXI-S), the pixel processing
electronics (SXI-PE), the digital electronics (SXI-DE), and the cooler
driver (SXI-CD). The SXI-S includes the CCDs and the single-stage
Stirling coolers (SXI-S-1ST) as well as analog electronics
for driving the CCDs (SXI-S-FE) and for processing the output signals
from the CCDs (Video board). The UserFPGA on the mission I/O board
in the SXI-PE generates CCD clock patterns and processes digitized
frame data from the Video board. The SXI-DE performs further data
processing and controls the heaters to keep the CCD temperature
constant. It also receives commands from the satellite management unit (SMU)
and sends telemetry including the frame and housekeeping data to the SMU
and the data recorder of the spacecraft.
The detailed design and specification of the SXI components
are summarized in another paper \citep{Tanaka17}.

%imaging system
Combining the SXT-I and the SXI, we realize an imaging system with a
focal length of 5.6~m and a field of view (FoV) of
\timeform{38'}~$\times$~\timeform{38'}.
This is the largest FoV among focal plane X-ray detectors
that cover up to 10~keV.
A large effective area and FoV yield a large collecting
power, $grasp = A\cdot\Omega_{eff}$, where $A$ is the effective
area and $\Omega_{eff}$ is the effective FoV. The soft X-ray
imaging system provides a grasp of 54~cm$^{2}$~deg$^{2}$ at 7~keV,
the largest among the focal plane X-ray CCD detectors \citep{Nakajima17}.
These characteristics not only work complementary to
the narrow-field, high resolution spectroscopy
of the Soft X-ray Spectrometer (SXS) \citep{Kelley17}
but also enable us to observe extended objects such as clusters of galaxies
and Galactic supernova remnants with a single pointing.

\section{Initial Operations}\label{sec:function}

Following the critical operations
such as the deployment of the solar paddles and the extension of 
the extensible optical bench, mission instruments were
sequentially started up.
Startup procedures of the SXI began 13 days after the launch
with the steps summarized in table~\ref{tab:ope}.
Function and performance checks were carried out during
the procedures as described in the following subsections.
The SXI switched to the regular event mode on 7th March after the initialization
of the dark level.

\begin{table*}[ht]
\caption{Startup procedures of the SXI.}
\begin{center}
\begin{tabular}{ll}
\hline\hline
Date in 2016 (UTC) & Operation \\
\hline
2nd March	&  SXI-DE and SXI-PE power on and initialization \\ 
3rd March	&  SXI-S-FE power on and initialization  \\ 
      	&  Load microcodes for inverse transfer mode \\
      	&  Activate autonomous commands for South Atlantic Anomaly (SAA)\\
4th March	&  Readout noise check with frame images \\
      	&  Load microcodes for observation modes \\
      	&  SXI-CD power on \\
5th March	&  SXI-S-1ST power on \\
6th March	&  SXI-S-1ST increase voltage \\
7th March	&  Initialize dark level and switch to the event mode \\
\hline
\end{tabular}
\end{center}
\label{tab:ope}
\end{table*}%

\subsection{Readout Noise}\label{ssec:ron}

Frame images were obtained just after the startup of
the front-end electronics. Because the CCDs were not yet
cooled, significant charges due to dark current were generated
in the sensors. Therefore we transfer the charges in the
serial register toward the opposite direction from the readout
node to measure the readout noise.
The standard deviations of the pulse height (PH) distributions
in the frame images are measured for all the segments.
The noise levels were distributed over a range of 4.3--5.3~ch
(analog-to-digital unit) for all the segments over 70 continuous frames.
While the scale between the PH and the incident X-ray energy depends on
the gain of each analog-to-digital converter (ADC),
1~ch approximately corresponds to 6~eV.
Parameters for the event detection and the dark
level calculation should be determined according to the readout
noise level. The values obtained in orbit agree with those measured
in the ground tests. Therefore we determined that
initial values of all the parameters such as split thresholds and
dark thresholds should be the same as those adopted in the ground tests.

\subsection{CCD Temperature}\label{ssec:temp}

The CCDs and a cold plate are cooled using one of the two SXI-S-1ST.
Two heaters are attached to the cold plate
and an application software in the SXI-DE controls the sensor temperature
by applying an electrical current to the heaters using a
proportional-integral-derivative controller \citep{Tanaka17}.
Considering possible accumulation of contaminants
including water onto the CCDs and consequent degradation of
quantum efficiency in the soft energy band, we set the operating
temperature to $-110~^{\circ}$C in the initial phase of the mission.
Because a charge transfer efficiency significantly
changes as a function of the CCD temperature \citep{2013NIMPA.731..160M},
it is crucial to keep the temperature constant during
observations for better spectroscopic performance.
The temperatures of all the CCDs were stable
within three digits corresponding to $0.24~^{\circ}$C in peak to peak
throughout the mission.

\subsection{Charge Injection}\label{ssec:ci}

We found that the charge transfer inefficiency (CTI) of the SXI CCDs is not
negligibly small in the ground calibration \citep{Tanaka17}.
Although we conjecture that the initial CTI is due to lattice
defects in the wafer or defects at the interface between the silicon substrate
and the oxide layer, the cause has not been identified.
Therefore the
charge injection (CI) capability was utilized since the beginning of the mission.
Although the structures of the injecting gate and drain are different from those
of the XIS \citep{2008PASJ...60S...1N}, we adopt the spaced-row CI method
as was applied to
the XIS \citep{2009PASJ...61S...1O, 2009PASJ...61S...9U}. The interval and
the offset of injected rows are 80 logical pixels and 1 in ACTY coordinate, respectively,
as those adopted in the ground tests (for ACT coordinate, see figure~7 in
\cite{Tanaka17}).
The amount of injected charge is set so that the PHs of the
injected pixels are saturated but that leakages to
the preceding and trailing rows do not affect the spectroscopic performance.

\begin{figure}[ht]
 \begin{minipage}{0.4\hsize}
  \begin{center}
%\vspace{2mm}
   \includegraphics[width=0.8\textwidth]{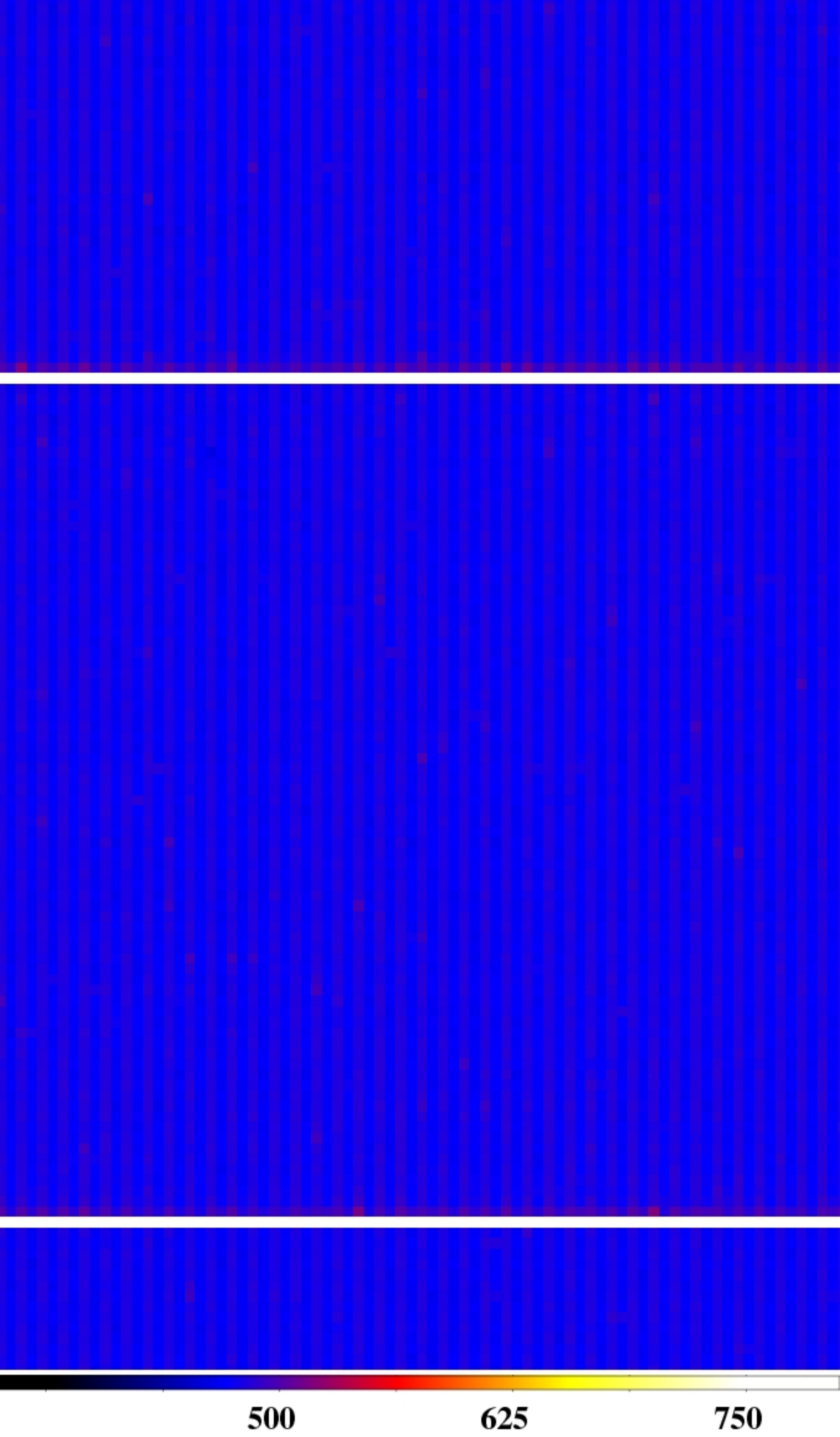}
  \end{center}
 \end{minipage}
\hspace{-3mm}
 \begin{minipage}{0.6\hsize}
  \begin{center}
   \includegraphics[width=1.37\textwidth]{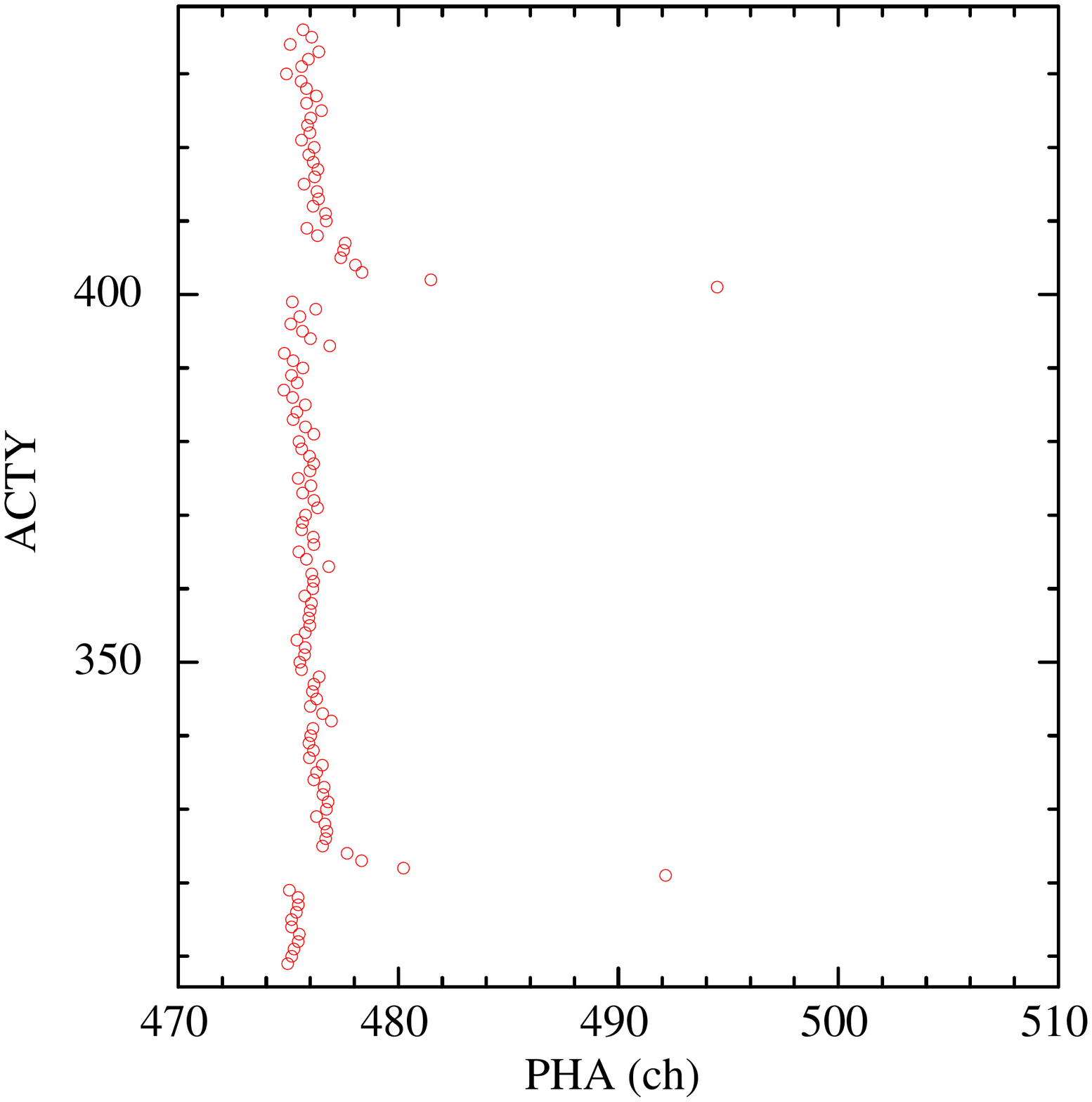}
  \end{center}
 \end{minipage}
\vspace{3mm}
 \caption{(Left) A part of dark frame image of CCD4 Segment~CD in the ACT coordinate
 obtained during the observation of the Perseus cluster of galaxies.
 Color scale indicates PH of each pixel.
 (Right) PH averaged for each row (ACTY) of the left image.
 Note that the PHs at the injected rows (ACTY = 321 and 401)
 are out of the range in abscissa.
 }\label{fig:chargeinjection}
\end{figure}

The PH distribution as a function of ACTY is investigated by measuring
the amount of the injected and leaked charges.
Figure~\ref{fig:chargeinjection}
shows an example of the dark frame image of CCD4 Segment~CD obtained after
the CCD temperature becomes stable. PHs averaged for each row of
the image are also shown. Note that the PHs of injected rows
(ACTY = 1 + 80~$\times$~n: n = 0--7) are
saturated and hence they are out of the range in abscissa in the right panel.
Only the first trailing rows (ACTY = 2 + 80~$\times$~n) exhibit
excess of PHs by approximately 20~ch compared with other rows.
The second trailing rows (ACTY = 3 + 80~$\times$~n) and the preceding rows
show a small amount of excess of $\le$5~ch. This value is certainly lower
than the split threshold of 15~ch that has been adopted since the ground tests.
Because an event is composed of 3~$\times$~3~pixels,
the PH of the event may not be
valid if a part of the 3~$\times$~3~pixels belongs to the injected
rows or the first trailing rows.
Then we filter out events in the standard screening
if their centers fall
in regions from the first preceding to the second trailing rows.

%\subsection{CBF (DYE)}
%new figure?

\section{Observations}
\label{sec:observation}  % \label{} allows reference to this section

%summary
Comprehensive function checks were followed by the scientific
observations and calibrations. Table~\ref{tab:obslog} summarizes
observation logs of all the celestial objects. Other pointings
for the purposes of checking the attitude and orbit control subsystem
(AOCS) are omitted.
All the observations were performed with the clocking mode of 
``Full Window + No Burst" except for the Crab nebula for which
the ``Full Window + 0.1-s Burst" mode was chosen for CCD1 and CCD2
to suppress pile-up.
Some parameters such as the event threshold and the area
discrimination were changed especially for the interval
when the minus Z axis of the spacecraft,
the direction from the mirror to the detector along the optical axis,
points to the day earth (MZDYE) as described in section~\ref{sec:issues}.
%notes for each pointing
The effective exposure times in table~\ref{tab:obslog} are
those excluding the MZDYE intervals.
Note that the pointing direction of the SXT-I
gradually shifted during the observation of N132D
\citep{1976ApJ...207..394D, Miller17},
a supernova remnant in the Large Magellanic Cloud, due to
unstable behavior of the AOCS.
%Therefore, the target resided within the SXI
%FoV with shorter duration than the effective exposure time.
The observation of IGR~J16318--4848 (\cite{2003IAUC.8063....3C};
Hitomi Collaboration 2017), a highly obscured high mass X-ray binary,
was performed
before optimizing the alignment matrices of the star trackers (STT),
which resulted in off-axis angles of \timeform{8'} and \timeform{5'} during the tracking
with STT1 and STT2, respectively. Its effective exposure time is the sum
of the two intervals.
%and the pointing direction is derived from the first one.
Although the SXS missed both targets during most of the exposure times,
we successfully obtained CCD spectra thanks to the large FoV of the SXI.

%software and CALDB used
All the event data are once processed using a script version 03.01.005.005
and subsequently reprocessed using the SXI calibration database
that reflects the ground calibration.
Results shown hereafter are obtained with the cleaned
event files unless otherwise noted.
The detailed criteria for the cleaning are summarized in the
calibration database
\citep{Angelini17}
\footnote{https:\slash\slash{}heasarc.gsfc.nasa.gov\slash{}docs\slash{}hitomi\slash{}calib\slash{}}.
In the spectral fit with
\texttt{XSPEC} \citep{1996ASPC..101...17A}, we adopt the model
\texttt{phabs} for the photoelectric absorption with
a solar metal composition of \citet{2000ApJ...542..914W}
and a photoionization cross-sections provided by \citet{1992ApJ...400..699B}.

\begin{table*}[ht]
%%\begin{landscape}
%\begin{longtable}{*{6}{l}}
\caption{Observation logs of the soft X-ray imaging system.} 
\label{tab:obslog}
\begin{center}       
%\begin{threeparttable} % 2017.4.11
%\begin{tabular}{llllll} %% this creates two columns
\begin{tabular}{llD{,}{,}{-1}D{.}{.}{-1}ll} %% this creates two columns
%% |l|l| to left justify each column entry
%% |c|c| to center each column entry
%% use of \rule[]{}{} below opens up each row
\hline\hline
Time$^\ast$ & Target & \multicolumn{1}{c}{\rm Pointing Direction} & \multicolumn{1}{c}{\rm Exposure} & Clocking Mode & Event Threshold$^\ddagger$ \\
                               &        & \multicolumn{1}{c}{\rm(RA, Dec) in  J2000} &\multicolumn{1}{c}{\rm (ks)}   & for CCD1 and CCD2$^\dagger$      & (ch)     \\
\hline
2016-03-06 22:55:00 & The Perseus cluster & (49.951,\ $+$41.512) & 40.3 & Full Window + No Burst & 40 \\
2016-03-08 00:38:00 & N132D & (81.246,\ $--$69.646) & 238.0 & Full Window + No Burst & 100 \\
2016-03-10 19:37:00 & IGR~J16318--4848 & (247.701,\ $--$48.832) & 142.1 & Full Window + No Burst & 100 \\
2016-03-16 19:40:00 & RX~J1856.5--3754 & (284.144,\ $--$37.909) & 11.4 & Full Window + No Burst & 40 \\
                                             &                  &   &    &        & 40(on-axis)/100(others)$^\S$\\
2016-03-19 17:00:00 & G21.5--0.9 & (278.388,\ $--$10.569) & 203.3 & Full Window + No Burst & 100$^\|$\\
2016-03-23 13:30:00 & RX~J1856.5--3754 & (284.145,\ $--$37.910) & 23.3 & Full Window + No Burst & 40(on-axis)/100(others)$^\#$ \\
2016-03-25 11:28:00 & Crab & (83.633,\ $+$22.013) & 0.242 & Full Window + 0.1-s Burst & 100$^\|$ \\
\hline
%\end{longtable} 
\end{tabular}
%\begin{tablenotes}\footnotesize
%\end{tablenotes}
%\end{threeparttable}
\end{center}
$\ast$ Time when the maneuver started to the target in the coordinated universal time (UTC).  \\
$\dagger$ Clocking mode for CCD3 and CCD4 is always set to ``Full Window + No Burst". \\
$\ddagger$ Note that 1~ch approximately corresponds to 6~eV.\\
$\S$ We changed the event threshold of all the segments except for the on-axis
one from 40~ch to 100~ch from 2016-03-18 18:45, while the event threshold is set to
2048~ch during MZDYE for all the segments throughout the pointing.\\
$\|$ During MZDYE, the event threshold is set to 100~ch and the area discrimination
(100 $\times$ 100~logical pixels) is applied to the on-axis segment, while the event threshold is set to
2048~ch and the area discrimination is off for other segments.\\
$\#$ During MZDYE, the event threshold is set to 40~ch and the area discrimination
is applied to the on-axis segment, while the event threshold is set to 2048~ch and the area discrimination is off for other segments.\\
\end{table*} 

\begin{figure*}[ht]
 \begin{minipage}{0.6\hsize}
  \begin{center}
   \includegraphics[width=0.9\textwidth]{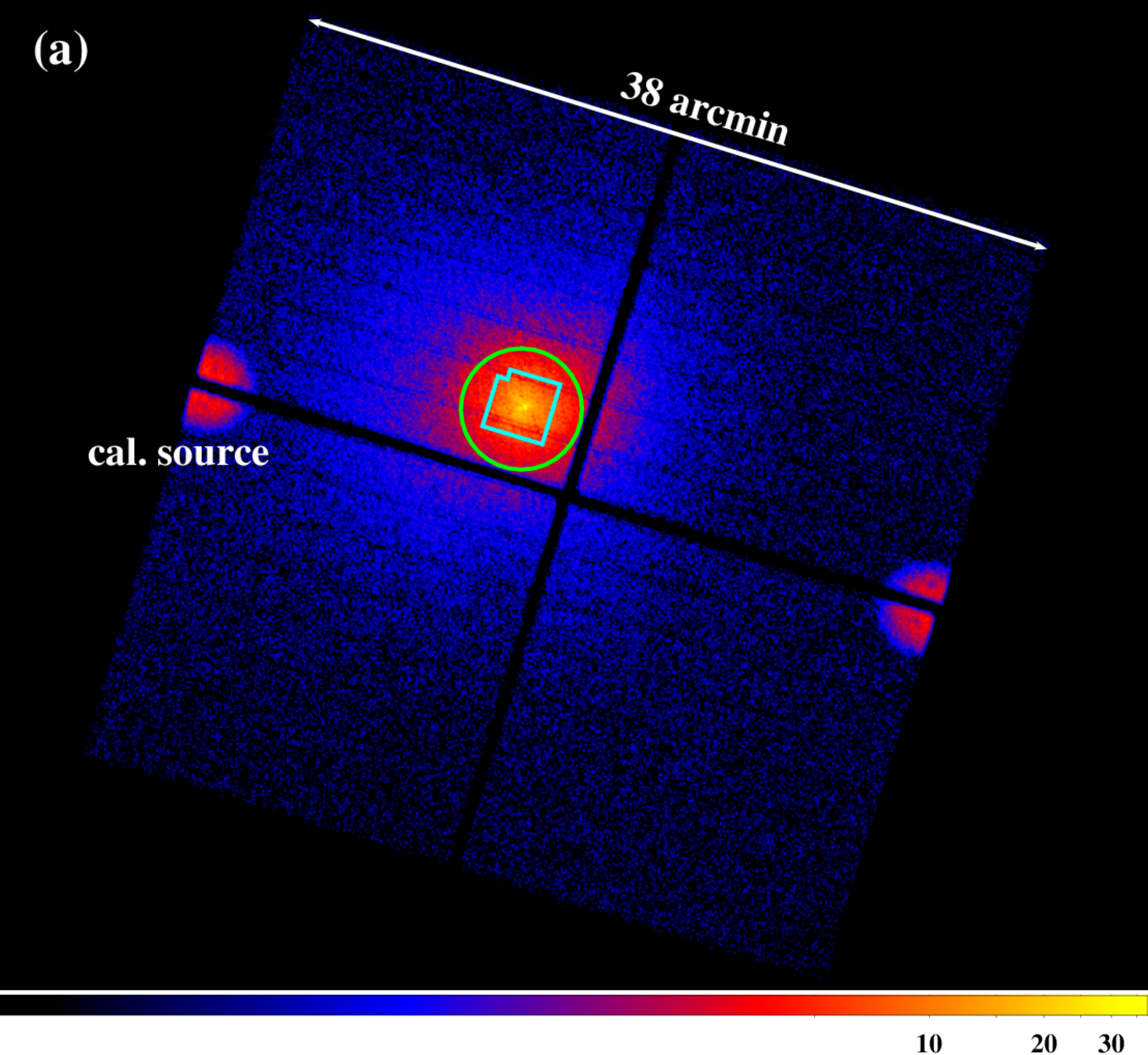}
  \end{center}
 \end{minipage}
 \hspace{5mm}
 \begin{minipage}{0.3\hsize}
  \begin{center}
   \includegraphics[width=\textwidth]{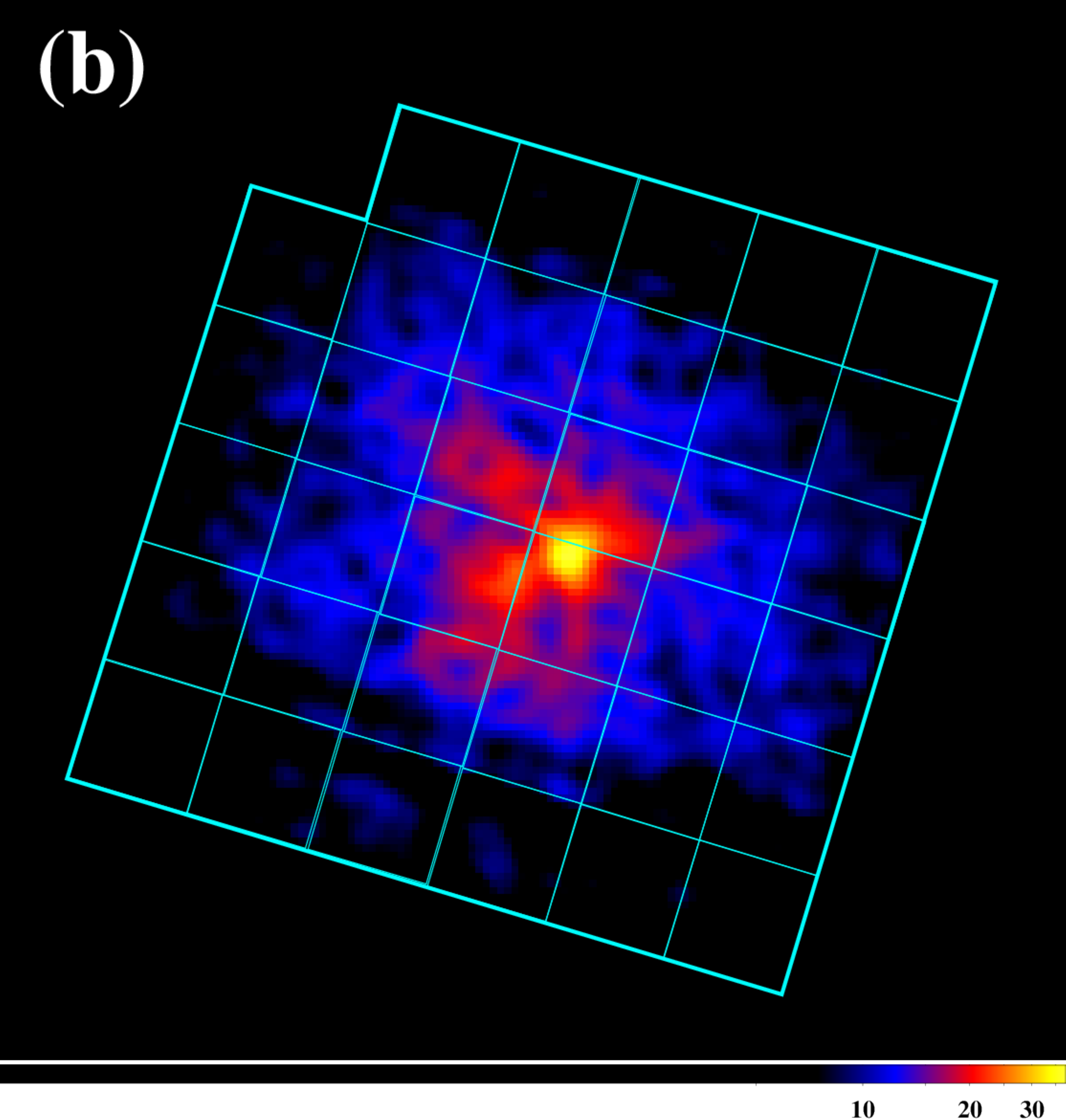}\\
   \includegraphics[width=\textwidth]{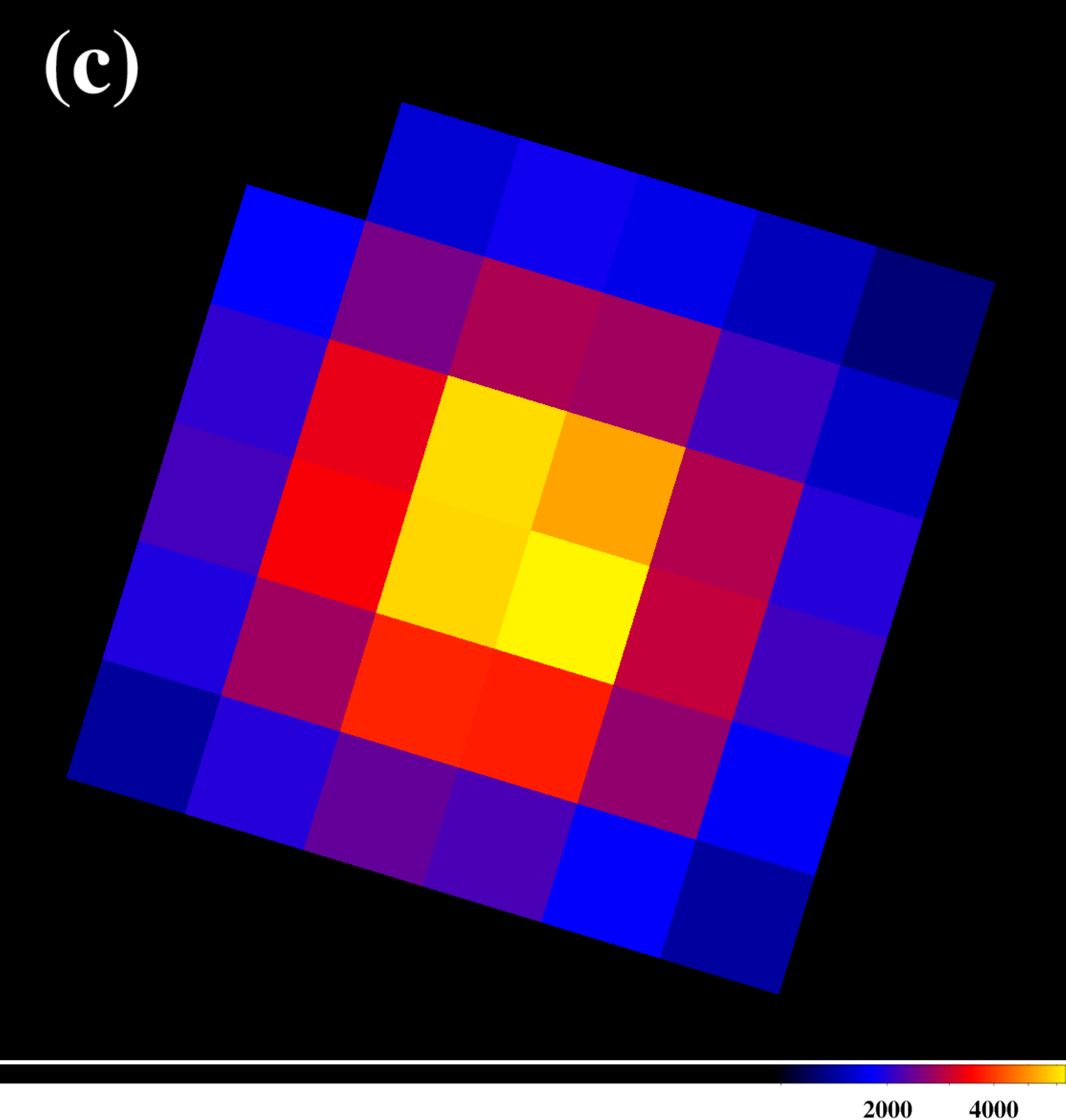}
  \end{center}
 \end{minipage}
 \vspace{5mm}
 \caption 
%>>>> use \label inside caption to get Fig. number with \ref{}
   { \label{fig:perseus} 
(a) SXI image of the Perseus cluster of galaxies in 0.4--12.0~keV
in the unit of counts per logical pixel
smoothed by a Gaussian of $\sigma$~=~3~pixels.
The bright spots at the edge of the FoV are due to the onboard calibration
sources. The cyan and green regions illustrate the SXS FoV and the photon extracting
region for figure~\ref{fig:Perseusspec}, respectively.
(b) Same as (a) but the entire FoV region of the SXS is extracted.
The regions of the detector pixels are overlaid for the comparison with (c).
(c) SXS image with the same energy band as (a) and (b).
}
\end{figure*}

%Perseus and its image
%The Perseus cluster, the first target for the soft X-ray imaging
%system, is the brightest cluster of galaxies in X-rays
%\cite{1992MNRAS.258..177E} and its
%emission extends over multiple CCD sensors. The core of the cluster
%is filled with the intracluster medium with an electron temperature
%of $\sim$4~keV. Central galaxy NGC~1275 hosts an active galactic
%nucleus. A notable quiescence of the hot gas with a line-of-sight
%velocity of 164$\pm$10~km/s is found in the region from 30 to 60~kpc
%from the central nucleus by the SXS \cite{2016Natur.535..117H}. 
Figure~\ref{fig:perseus} shows the first light SXI image of
the Perseus cluster of galaxies.
We confirm that the FoV of the soft X-ray imaging system
covers the central region of the Perseus cluster of galaxies
with a diameter of approximately 650~kpc.
Events originating from two $^{55}$Fe onboard calibration sources
are seen at the edges of the FoV.
The offset of the center of the four CCDs with regard to the center
of SXS FoV is found to be slightly smaller than the design value,
making the center of the SXS FoV nearer to the gaps of CCDs by \timeform{65''}.
We also compare the images obtained with the SXI and the SXS.
The size of a CCD logical pixel is 48~$\mu$m that
corresponds to \timeform{1.768''}. Thanks to the smaller plate scale
compared with that of the SXS (\timeform{29.982''}/pixel), the position and the
flux of the point source can be precisely measured.

%Perseus Image

\begin{figure}[ht]
 \begin{center}
  \includegraphics[width=\linewidth]{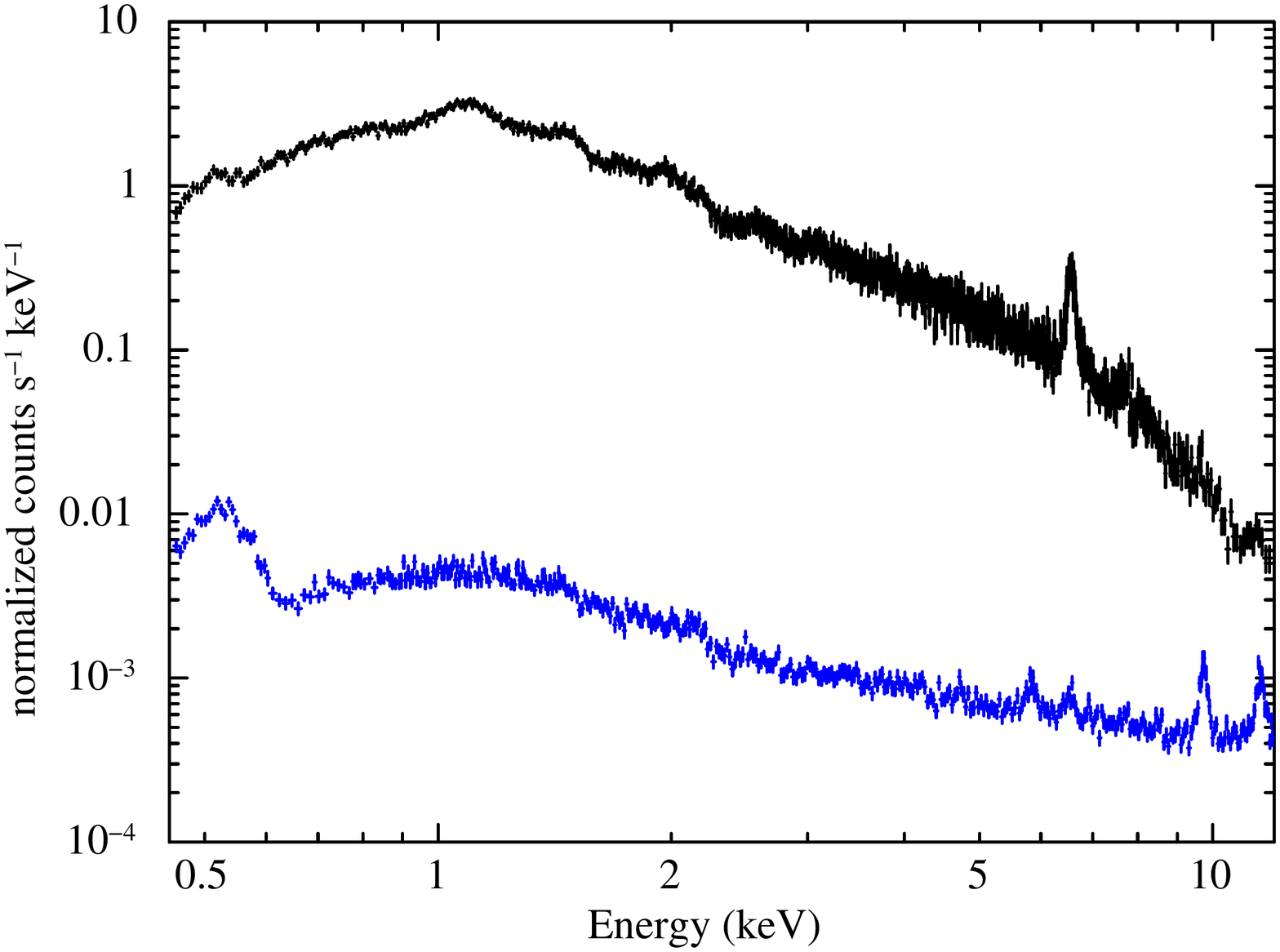}
	\begin{picture}(0,0)
	\put(-75,85){\includegraphics[height=2.1cm]{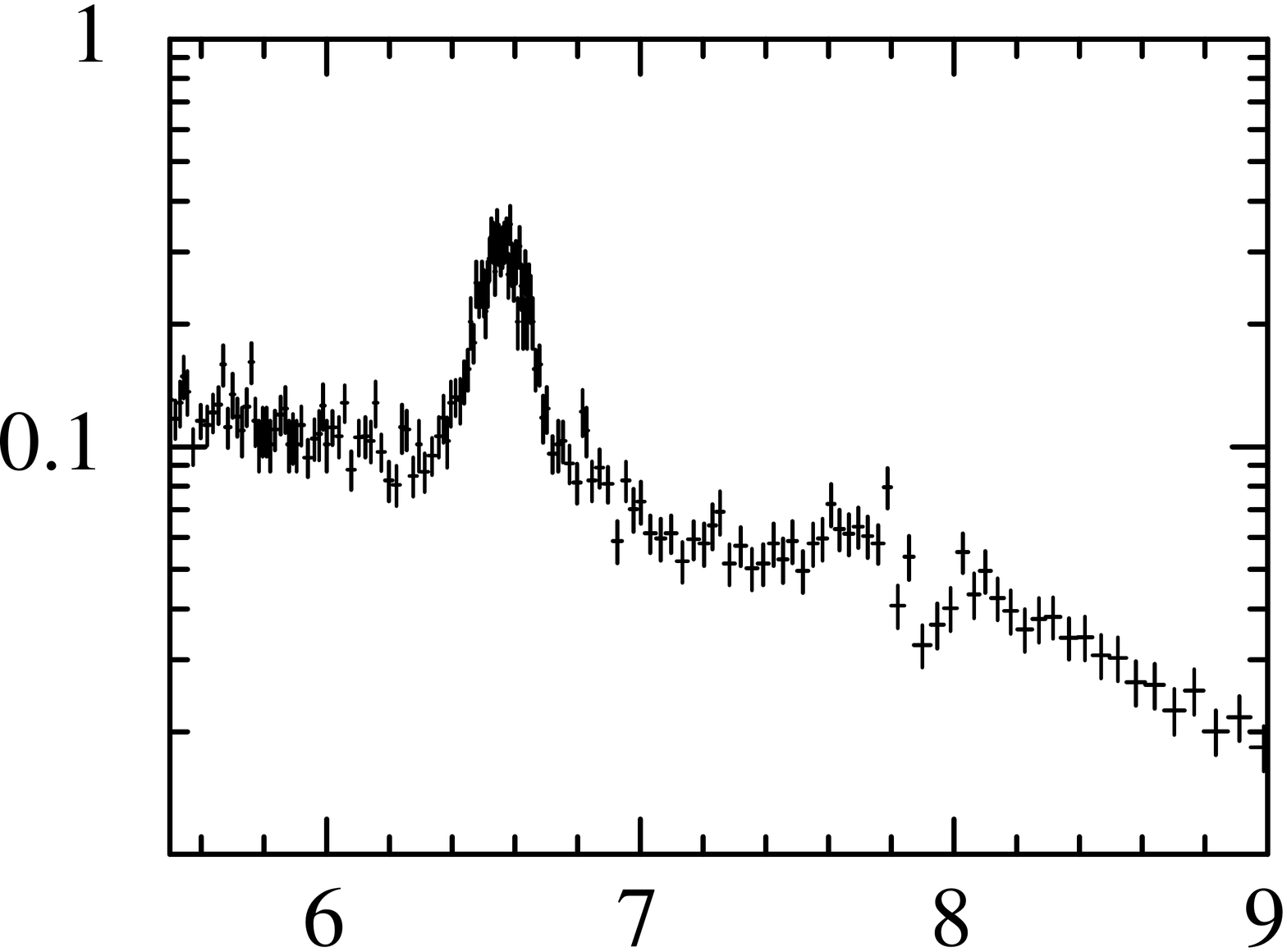}}
%		\put(-150,250){\includegraphics[height=5cm, angle=-90]{src-r3arcmin_bgd-wholeFoV-r12min_perseus_inset.ps}}
	\end{picture}
 \end{center}
 \caption{SXI source and background spectra of the Perseus cluster of galaxies
 in the energy band of 0.45--12.0~keV.
 The source spectrum (black) is extracted
 from the circular region with a radius of \timeform{3'} as shown in figure~\ref{fig:perseus},
 and the background spectrum (blue)
 is extracted from the entire FoV excluding a circular region with a radius of
 \timeform{12'} and the same center as the source region. The calibration source regions
 are also excluded. The inset shows the zoom-up view of the source spectrum
 near the Fe He$\alpha$ and He$\beta$ lines.
 }\label{fig:Perseusspec}
\end{figure}

The source and background spectra of the Perseus cluster are shown in
figure~\ref{fig:Perseusspec}.
The source spectrum is extracted from the core region with a radius of \timeform{3'}
as shown in figure~\ref{fig:perseus}, while the background spectrum comes from
the entire FoV excluding the core region with a radius of \timeform{12'}
and the calibration source regions.
The strong emission line of Fe He$\alpha$ (6.7~keV, rest frame), the weak lines of
Fe Ly$\alpha$ (7.0~keV), and the mixture of Fe He$\beta$
and Ni He$\alpha$ (7.8--7.9~keV) can be seen, which demonstrates
the spectroscopic performance of the soft X-ray imaging system.

\section{Issues Found in Orbit}\label{sec:issues}

%We are confronted with two issues after switching to the event mode:
%light leak and cross talk.
A part of the data are affected in terms of the effective exposure time
and the effective area by two issues: a light leak and a cross talk.
Following subsections elaborate on the problems and the measures.

\subsection{Light Leak}\label{ssec:LL}

\subsubsection{Light Leak during MZDYE}\label{sssec:IntenseLL}

The ground tests using a visible light LED revealed the light
leak through pinholes in the optical blocking layer (OBL)
on the CCDs \citep{Tanaka17}.
The regions near the physical edges of the CCDs were also found to be
sensitive to visible light.
%and that in the regions near their physical edges.
Then we thickened Al layers in the contamination blocking filter (CBF),
which was originally implemented to block contaminants onto the CCDs,
to accommodate the requirement regarding the transmittance of
the visible light and ultraviolet.
The CBF was confirmed to have survived the launch environment 
with data when the plus Z axis of the spacecraft points to the day earth.
However, it was found that visible light and/or infrared unexpectedly
entered the sensor during MZDYE through holes in the base panel of
the spacecraft. These incoming lights induced considerable number
of false events at the positions of pinholes
and the physical edges identified in the ground tests.

\begin{figure}[ht]
 \begin{center}
\vspace{-20mm}
  \includegraphics[width=1.2\linewidth, bb=50 0 842 595]{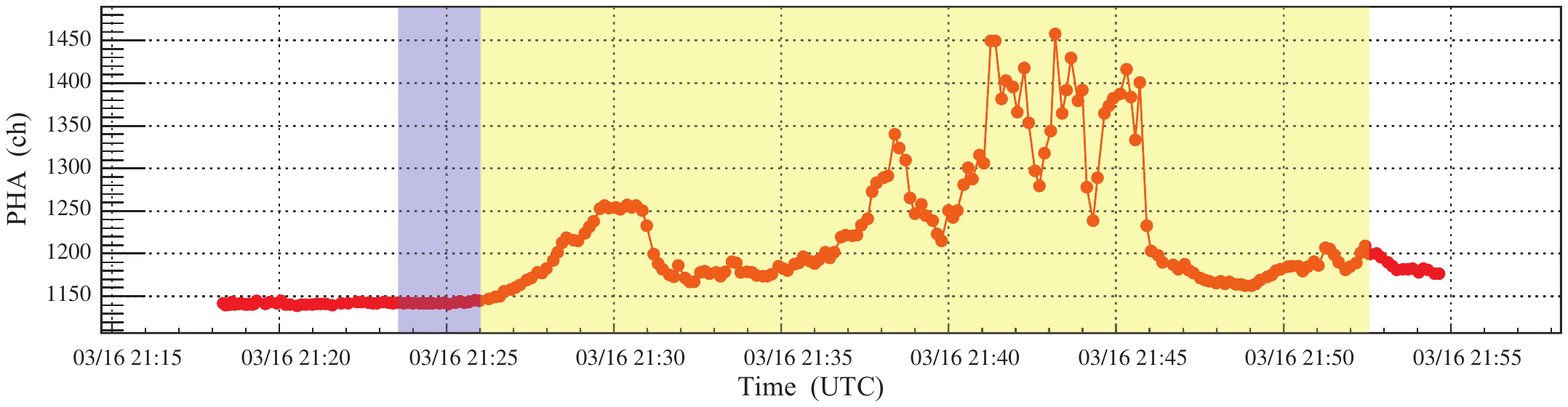}
\vspace{-20mm}
 \end{center}
 \caption{Example history of the PH of the light leak pixels.
 All the affected pixels in the region of ACTX = 4--316 and ACTY = 7--550 in
 CCD4 Segment~AB are averaged. Intervals when
 the minus Z axis points to the day and night earth are hatched in yellow
 and blue, respectively.
 }\label{fig:lightleak_lightcurve}
\end{figure}

We define the light leak pixels as those whose
median PHs during MZDYE exceed those during
non-MZDYE by more than 40~ch. Then approximately 7\% of the pixels
in the region that corresponds to the SXS FoV are
classified as the light leak ones.
Figure~\ref{fig:lightleak_lightcurve} shows an example of the PH
history averaged among the light leak pixels in CCD4 Segment~AB
that is a half region of the imaging area of CCD4 (for the segment designation,
see figure~4 in \cite{Tanaka17})
during the observation of the isolated neutron star
RX~J1856.5--3754 \citep{1996Natur.379..233W}.
%Even though no transient source appeared in this observation,
The PH began to increase at the start of MZDYE
and became highly time-variable during MZDYE.
Because the event threshold was set to 40~ch in this period,
a large number of events are produced due to this light leak.
%light path
%There are holes
%in the base plate of the spacecraft to ensure the light paths of the
%optics for Canadian Astro-H Metrology System \cite{2014SPIE.9144E..56G}
%and the hard X-ray imaging system \cite{2014ApOpt..53.7664A, 2016NIMPA.831..235S}.
%The incoming lights are thought to be reflected by the bottom surface of the lower
%plate and then pass through the contamination blocking filter (CBF).
%number of pixels affected
%The time profiles of the event flares differ from each other,
%which may reflect the sun angle and albedo at the surface of the earth
%where SXI look at through the holes in the base plate.
%The time variability of the PH is extensively investigated
%for all pixels using frame data.

%couter measure
To eliminate these false events, we defined a new
screening criterion to produce scientifically useful good time intervals.
When the minus Z axis of the spacecraft points to the earth and
an elevation angle of the minus Z axis above the
day earth limb is larger than that above the night earth limb, there is
little effect due to the light leak because the boundary of the day earth
and the night earth is clear. On the other hand, the boundary between
the day earth and the sky is indistinct in terms of the light flux
due to the earth atmosphere. In fact, the average PH changes
even after MZDYE as seen in figure~\ref{fig:lightleak_lightcurve}.
Considering these circumstances, we define the screening criterion
using the extended housekeeping data as \\
\texttt{
%\begin{equation}
(MZDYE\_ELV\ >\ MZNTE\_ELV)\ ||\ (MZDYE\_ELV\ >\ 20),\\
%\end{equation}
}
where
%MZNTE means the pointing of the minus Z axis 
%of the spacecraft toward the night earth. Note that
\texttt{MZDYE\_ELV} and \texttt{MZNTE\_ELV}
mean the elevation angles of the minus Z axis of the spacecraft
from the day and night earth limbs, respectively, in the unit of degree.

\subsubsection{Light Leak during the Sun Illumination of the Spacecraft}\label{sssec:MildLL}

\begin{figure}[ht]
 \begin{center}
  \includegraphics[width=\linewidth]{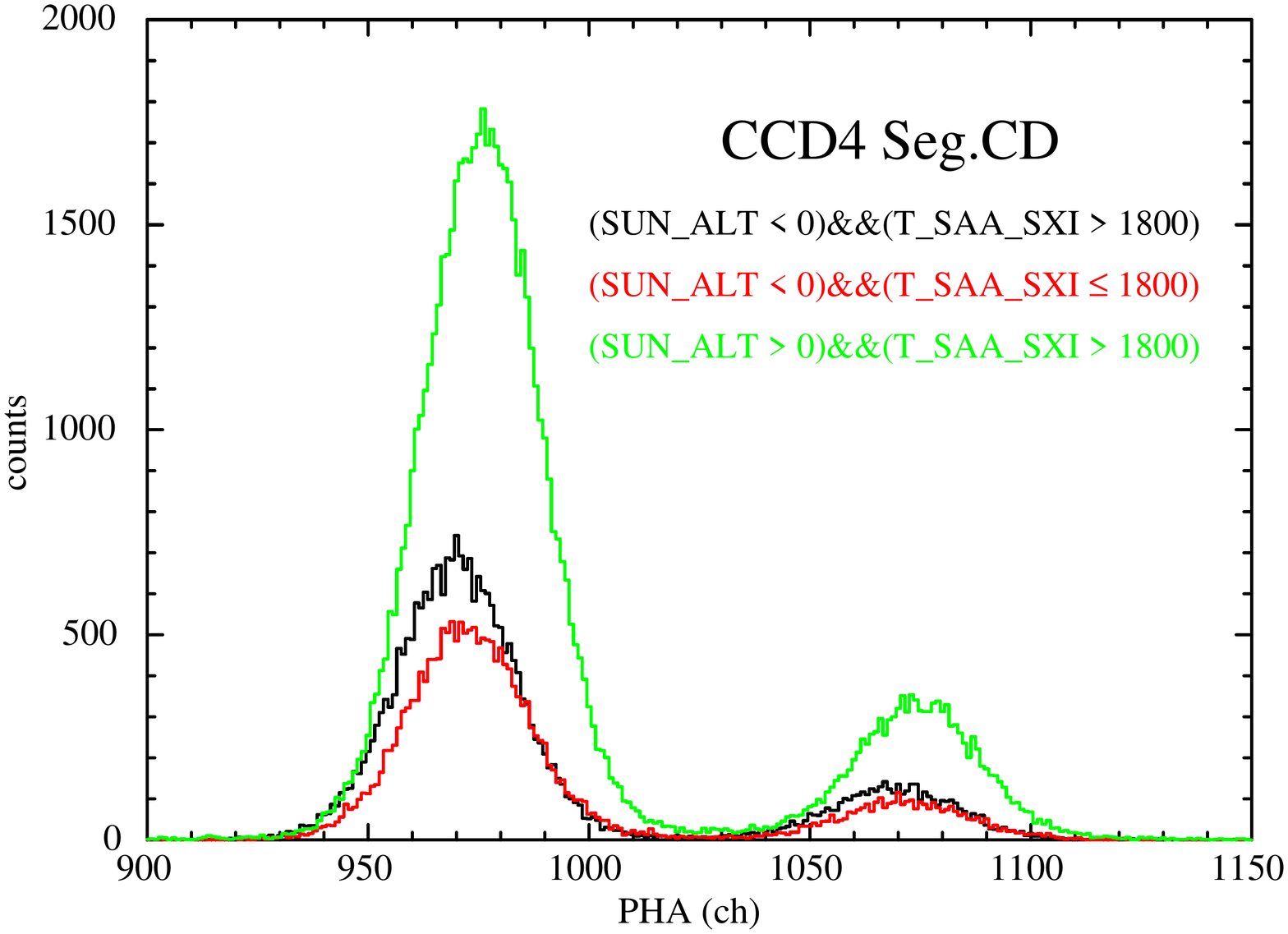}
 \end{center}
 \caption{Spectra of the calibration source events obtained with CCD4 Segment~CD.
 Spectrum integrated during the eclipse of the spacecraft (\texttt{SUN\_ALT}~$<$~0)
 and 1800~s or later after the passage of the SAA (\texttt{T\_SAA\_SXI}~$>$~1800)
 is shown in black. Data during the eclipse but \texttt{T\_SAA\_SXI}~$\le$~1800 are shown
 in red, while those during the sun illumination of the spacecraft
 (\texttt{SUN\_ALT}~$>$~0) and \texttt{T\_SAA\_SXI}~$>$~1800 are shown in green.
 }\label{fig:55fe_sunshine_sunshade}
\end{figure}

%sunshine and sunshade
Even though we can eliminate the light leak effect during MZDYE with
the above screening criterion, there is another small effect seen during
the sun illumination of the spacecraft.
Figure~\ref{fig:55fe_sunshine_sunshade} shows the spectra of the
Mn K lines from the onboard $^{55}$Fe calibration source detected
by CCD4 Segment~CD. Note that only the gain difference between two
ADCs in the Video board \citep{2013NIMPA.731..166N} is corrected
while the charge trail and CTI corrections are not performed.
We utilize \texttt{SUN\_ALT}, the solar altitude from the earth limb,
in the extended housekeeping data to discriminate the intervals of the
sun illumination of the spacecraft (\texttt{SUN\_ALT}~$>$~0) and
the eclipse of the spacecraft (\texttt{SUN\_ALT}~$<$~0).
For further investigation of a possible shift of the line centroid depending on the
time after the passage of the SAA (\texttt{T\_SAA\_SXI}) in the unit of second,
we divide the data during the eclipse of the spacecraft
into two with the boundary of \texttt{T\_SAA\_SXI}~=~1800.
The difference of the line centroids between the eclipse and sun illumination
with \texttt{T\_SAA\_SXI}~$>$~1800
is 5.8~$\pm$~0.1~ch that corresponds to approximately 35~eV.
We see this effect because the region irradiated
with the calibration sources is close to the physical edge of the CCDs
that is most sensitive to the visible light.
N132D and IGR~J16318--4848 were unexpectedly caught at the edge of CCD1
and hence an additional screening with regard to \texttt{SUN\_ALT}
may be needed after the standard screening.

%T_SAA
There is also a small difference of the line centroid between
the spectra summed during \texttt{T\_SAA\_SXI}~$\le$~1800
and \texttt{T\_SAA\_SXI}~$>$~1800.
There may be some charges left at least near the physical edges
of the CCDs after the passage of the SAA.
We stop the sequence of driving and reading out the CCDs during the SAA.
%Furthermore, every voltage applied to the CCDs are switched to the lowest values.
Because this operation leads to a considerable amount of
charges accumulated in the imaging area, we sweep up the charges
just after the passage of the SAA by starting the sequence with the reversed
electric field in the substrate. One of the possible reasons of
the line shift depending on the time after the passage of the SAA
is that the parameters used to sweep up
the charges such as the applied voltages and the duration of the sweeping
are not optimized.

\subsection{Cross Talk}

\begin{figure}
  \begin{center}
  \includegraphics[width=0.8\linewidth]{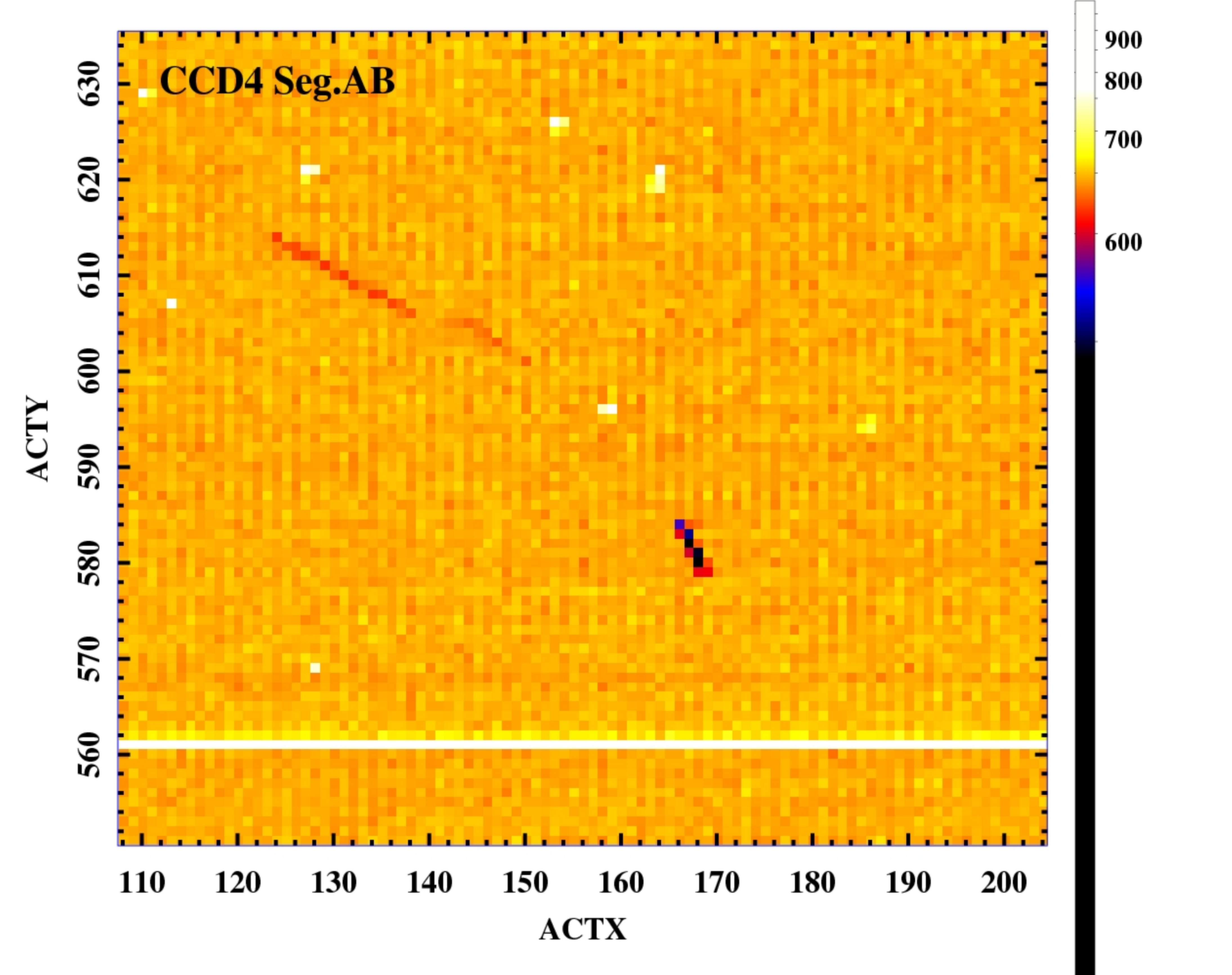}\\
  \vspace{2mm}
  \includegraphics[width=0.8\linewidth]{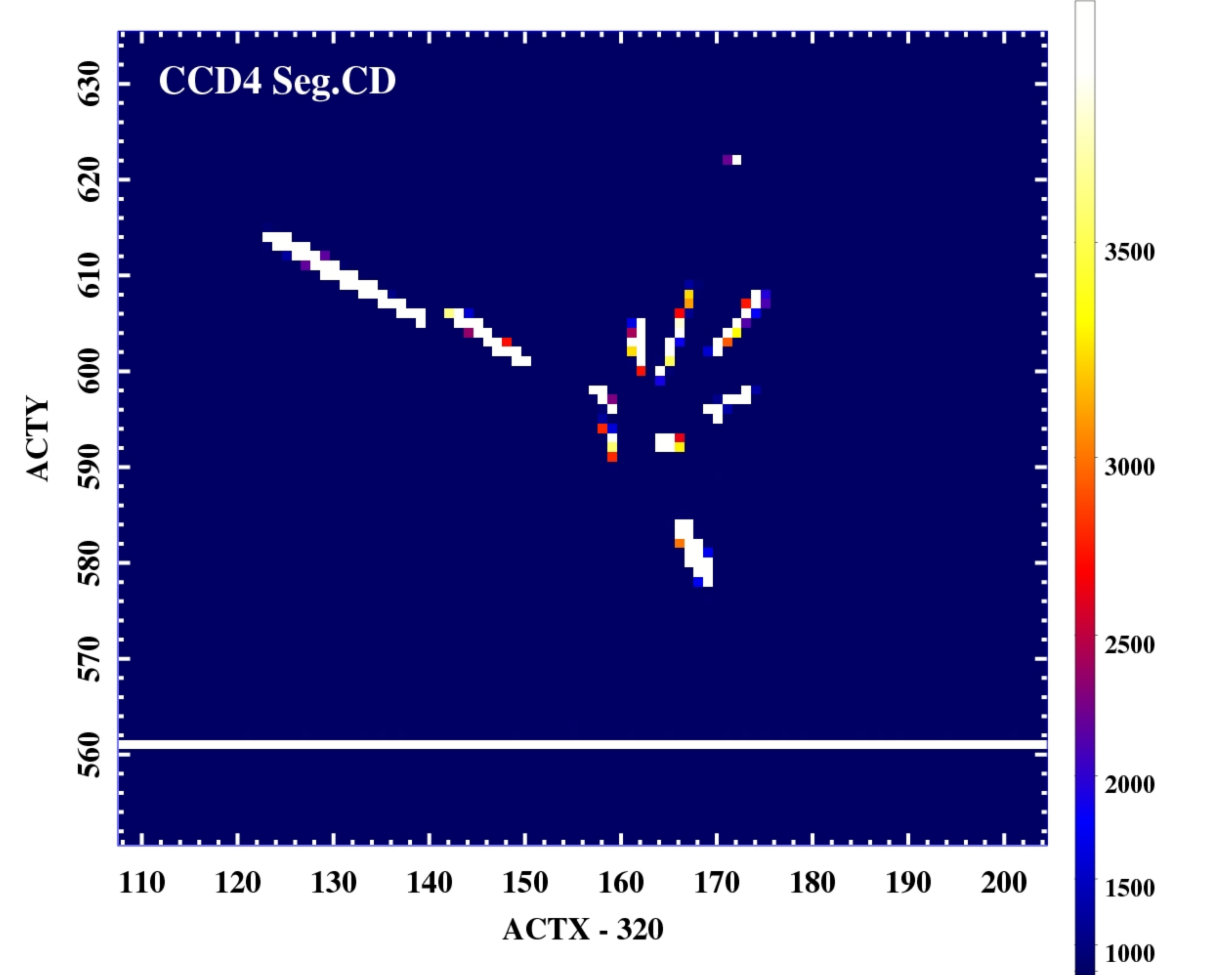}
  \end{center}
 \caption{Zoom-up view of raw frame images obtained with 4~s exposure
 for the two segments of CCD4
 obtained at the same time in orbit. The large number of charges distribute
 along the tracks
 of the charged particles in the Segment~CD. The pixels
 in the Segment~AB read out at the same time as those in the Segment~CD
 with high PHs exhibit lower PHs compared with those of surrounding pixels.
 }\label{fig:crosstalk_image}
\end{figure}

%phenomenon
An extraordinary large signal from a pixel may cause a cross talk with
negative polarity to induce an anomalously low PH in the pixel
at the same coordinate of the adjacent segment (e.g., \cite{2012PASP..124.1347Y}).
Such cross talk produced false events which occupied the telemetry
during the initial phase.
Figure~\ref{fig:crosstalk_image} shows raw frame images from the same
part of the adjacent segments obtained at the same time. Charged particles
produce a large amount of signal charges along their tracks in the CCDs
and also produce
negative signals in the adjacent segment. The PHs of the pixels
return to their normal levels in the next frame. Nevertheless, if the negative
PHs fall below the lower threshold for the dark level update,
the anomalously low dark levels are set and kept for the pixels
because the original PHs are recognized to be high
due to signals by X-rays or charged particles.
In this way, the cross talk produces false events continuously from these pixels.
Because the particle
events occur randomly in time and position, the cross talk results in the
gradual increase of the event rate.
The PHs of the false events are mostly below 100~ch in this case.
Also, when the affected pixels are in the 3~$\times$~3 pixels of
the normal events, the PHs of the events become higher than the
true values by similar amounts of the PHs of the false events.

\begin{figure}
 \begin{center}
  \includegraphics[width=\linewidth]{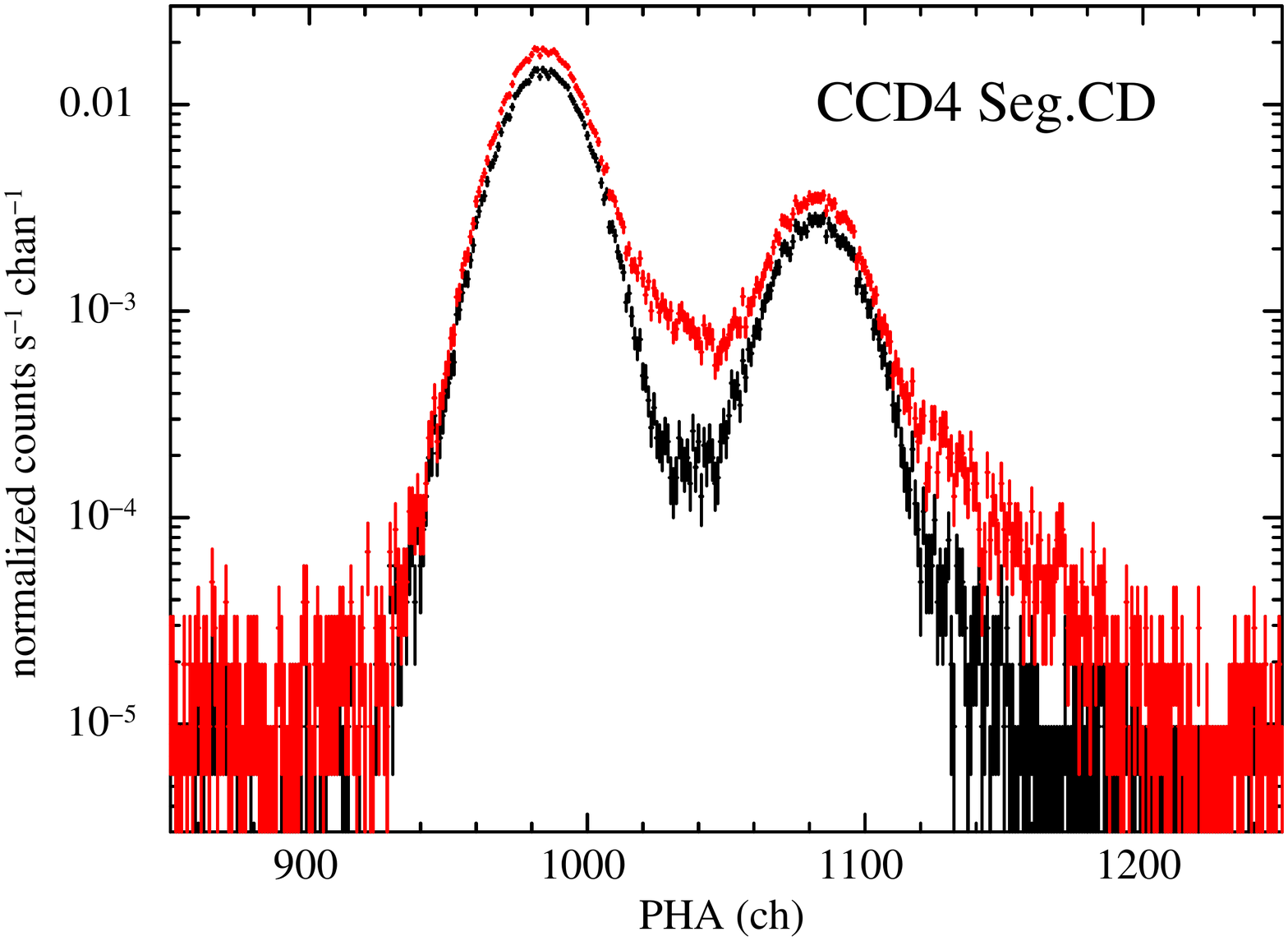}
 \end{center}
 \caption{Mn K spectra during the first half of the observation
 of IGR~J16318--4848. Filtering of the affected pixels by cross talk issue
 is applied to the data in black but not for the data in red.
 Parameters in the filtering are $N_{min}$ of 6, $PHA_{sp}$ of 15, and $R$ of 0.7.
 }\label{fig:crosstalk_55Fe}
\end{figure}

%countermeasure
The effect of the cross talk would not have appeared if we had optimized the threshold
for the dark level update so that the negative PHs fell securely
within the thresholds. Because the dark level cannot be changed retroactively,
we introduce an additional screening for the cross talk issue
in the software \texttt{sxipipeline}.
If at least a fraction ($R$) of the events containing a pixel of interest
have a PH greater than or equal to a split threshold ($PHA_{sp}$)
in that pixel, we consider that the pixel is affected
by the cross talk and eliminate it from further analyses.
This judgement is applied
to all the pixels that is contained in at least $N_{min}$ events.
%For a pixel that is contained in 5~$\times$~5~pixels for at least
%a specific number of events ($N_{min}$), if its PH is above a split
%threshold ($PHA_{sp}$) for at least a fraction ($R$) of those events,
%then the pixel is considered to be affected by the cross talk and is discarded.
$N_{min}$, $PHA_{sp}$, and $R$ can be adjusted by users.
The effect of this additional screening is illustrated
with spectra of Mn K lines in figure~\ref{fig:crosstalk_55Fe}.
Because the inappropriate dark levels tend to be slightly lower than
the normal values, the PHs of the affected events are slightly higher than
those of the normal events. This effect induces the high-energy tails that
are substantially decreased after the screening.
%the enhancement in the higher outskirts of each line.
%After we apply the additional screening with $N_{min}$ of 6,
%$PHA_{sp}$ of 15, and $R$ of 0.7,
%the outskirts in the Mn K spectra notably decrease.

%on-axis segment
Curiously, the polarity of the cross talk signal is
opposite only in the on-axis segment, possibly due to the difference
in capacitive and/or inductive coupling along the signal path
from the CCDs to the ASICs. Thus the on-axis segment does not
suffer from the false events. According to an exposure map
summed over the intervals when the spacecraft point to the night
earth, the exposure time of the off-axis segments decrease by approximately
17\% of that of the on-axis segment.
% as shown in the right panel
%in Fig.~\ref{fig:crosstalk_lightcurve}.
%Figure~\ref{fig:NXB_exposure}
%shows the exposure map summed over the duration when the satellite
%point the night earth to integrate non-X-ray background (NXB) data.
%Summation of the exposure is performed after eliminating the affected
%pixels by cross talk for each event file. While the exposure times
%in the on-axis segment is uniform, those in other segments differ
%pixel by pixel depending on when the cross talk happened in the pixel.

%\begin{figure}
% \begin{center}
%  \includegraphics[width=\linewidth]{NXB_expmap_55Fe.eps}
% \end{center}
% \caption{Exposure map of the NXB data in DET coordinate with a unit of seconds.
% The pixels affected by the cross talk is filtered for each event file and
% then the exposure maps are integrated. Parameters for the filtering are
% the same as those adopted in Fig.~\ref{fig:crosstalk_55Fe}.
% }\label{fig:NXB_exposure}
%\end{figure}

\section{Calibration}\label{sec:cal}

\subsection{Energy Scale and Energy Response}\label{ssec:gainCTI}

%gain
The energy scale
of the SXI is monitored with Mn K$\alpha$ and K$\beta$
from the calibration sources for all the CCDs.
To minimize the effects described in section~\ref{ssec:LL},
we extract the data during the eclipse of the spacecraft
and \texttt{T\_SAA\_SXI}~$>$~1800.
%The history of the line center when we fit the Mn K$\alpha$ line
%with a single Gaussian model is shown in Fig.~\ref{fig:55fe_trend}.
The mean PHs show significant fluctuations within the range of $\sim$~2--3~ch,
which corresponds to $\sim$~12--18~eV. This can be interpreted
as a systematic uncertainty on the SXI energy scale.
%When the trend is fitted with a linear function, there is no significant
%change in the line center during the three weeks. 
%\begin{figure}
% \begin{center}
%  \includegraphics[width=1.0\linewidth,bb=0 0 667 480]{55Fe_trend_nte.png}
% \end{center}
% \caption{Trend of the pulse height of the Mn K$\alpha$ line center.
% Each data point corresponds to a observational sequence.
% {\bf THE FILES WILL BE UPDATED IN FUTURE.}
% }\label{fig:55fe_trend}
%\end{figure}
%FWHM
We also fit the spectra of the calibration sources with a response
function determined from the ground calibration \citep{2016NIMPA.831..415I}.
%Energy resolution is measured by introducing an additional line width
%as a free parameter.
%%The history of the additional widths
%%is shown in Fig.~\ref{fig:55fe_fwhm_trend}.
Almost all the data require an extra line width even we extract data only during
the eclipse of the spacecraft.
Here we define the energy resolution to be the full width at
half maximum (FWHM) of the primary Gaussian component in the response function
\citep{2016NIMPA.831..415I}.
%The additional widths are 17.0, 22.0, 24.3, and 26.9~eV for CCD1, 2, 3, and 4,
%respectively, and they are constant as a function of time.
%Considering these additional widths,
The energy resolution averaged among
all the segments is 179~$\pm$~3~eV at 5.89~keV, 
which is larger than the value determined from the ground calibration
test by $\sim$~10~eV \citep{Tanaka17}. Although the reason of the
degradation is not clear, there might be remaining effect by the
issues described in section~\ref{sec:issues}.
%Fitting the history with a linear function, the slope is negligible
%for all sensors, indicating that the degradation is not time dependent.
All the spectral analyses
shown hereafter are performed using the updated response function.

%\begin{figure}
% \begin{center}
%  \includegraphics[width=0.9\linewidth]{CCDall_sighist_dev_rev170530.eps}
% \end{center}
% \caption{Trend of the additional widths needed in the fit of the Mn K$\alpha$
% lines with the response function determined in a ground calibration.
% Diffrent colors designate CCD ID (CCD1:black, CCD2:red, CCD3:green, CCD4:blue).
% }\label{fig:55fe_fwhm_trend}
%\end{figure}

\begin{figure}
 \begin{center}
  \includegraphics[width=\linewidth]{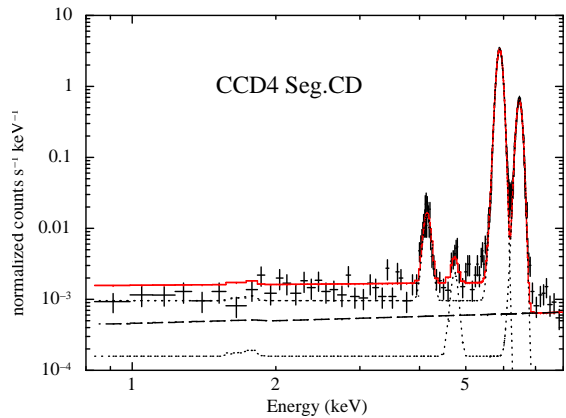}
 \end{center}
 \caption{Mn K spectrum of CCD4 Segment~CD accumulated during the observation
 of G21.5--0.9. The events are extracted in the period of the eclipse of the spacecraft
 and of 1800~s or later after the passage of the SAA.
 The spectral model shown with a dashed line is a power-law model to reproduce
 NXB component.
 }\label{fig:55fe_inorbit}
\end{figure}

%Response
The line profile in the response function is calibrated using
the Mn K spectrum integrated over long duration.
Figure~\ref{fig:55fe_inorbit} shows a spectrum
from one of the $^{55}$Fe calibration sources obtained
during the observation of the pulsar wind nebula G21.5--0.9
(\cite{2010ApJ...724..572M}; Hitomi Collaboration in preparation).
%Besides the standard screening, we only extract the events that
%detected during the sunshade of the spacecraft and the time
%after the passage of SAA of larger than 30~minutes.
The exposure time is 76~ks after the screening with
\texttt{SUN\_ALT} and \texttt{T\_SAA\_SXI}.
The Mn K$\alpha$ and K$\beta$ lines are fitted with Gaussian functions
using the updated response function.
We add a power-law model to represent the non-X-ray background (NXB).
The intensities of the Si escape lines and the constant components
are well reproduced with the line profile parameters determined
from the ground calibration \citep{2016NIMPA.831..415I}
in which we utilized fluorescence lines and monochromatic X-rays.

\subsection{Effective Area and Quantum Efficiency}\label{ssec:QEandEA}

In-flight calibration of the quantum efficiency and the effective area
was carried out with the data of the Crab nebula.
%(catalogue designations M1, NGC~1952, Taurus A, SN~1054) . 
At the center of the nebula, the Crab pulsar lies with a spin period
of $\sim$~34~ms and is viewed as a point-like source. 
In the X-ray band, the Crab nebula including its central pulsar emits
synchrotron radiation which is reproduced by an absorbed power-law model. 
%Its phase averaged flux is fairly persistent.
%The Crab nebula has been used for an effective area calibration as
%a standard candle
\citet{1974AJ.....79..995T} suggested that the 2--10~keV flux shows 15\%
variation. \citet{2011ApJ...727L..40W} also reported that the flux in the
2--15~keV band varies by approximately 10\% over long timescales.
While the variability limits the reliability of the flux estimation
to the level of approximately 10\%, the Crab nebula has been used
as a standard candle for effective area calibration
\citep{1974AJ.....79..995T,2005SPIE.5898...22K,
2010ApJ...713..912W,2015ApJ...801...66M}.

%The SXT-I is aligned to image an on-axis source at the nominal position
%that is located near the edge of the CCD2. 
Since the Crab nebula is as bright as $\sim$~3~$\times$~10$^3$~counts~s$^{-1}$~SXI$^{-1}$
at the nominal position, it is important to minimize pile-up. 
The ``Full Window + 0.1-s Burst" mode was then adopted for CCD1 and CCD2. 
This mode provides the shortest exposure per frame and hence
reduces the pile-up probability as much as possible \citep{Tanaka17}. 
Its integration time per 1 frame cycle is as short as 0.0606~s. 
CCD3 and CCD4 were operated with the nominal exposure of 4~s. 
The data are screened with the standard criteria 
and further screened for intervals in which the pointing is expected to be stable.

\begin{figure}
  \hspace{15mm}
\vspace{-20mm}
 \begin{center}
  \includegraphics[width=\linewidth]{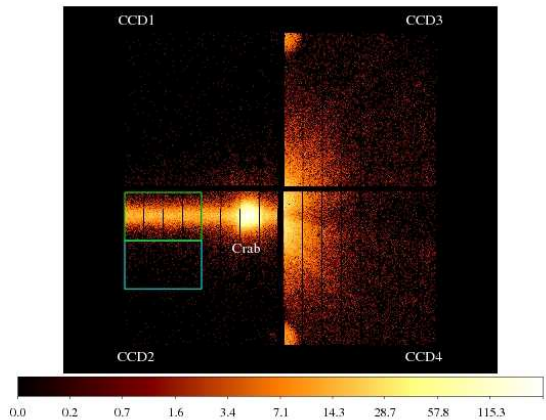}
  \end{center}
\vspace{-25mm}
  \caption{SXI image of the Crab nebula observation in the DET coordinate with no
  energy cut. The source and background events are extracted from green and
  cyan regions, respectively. The size of the extracted region is
  320~$\times$~200~pixels for both of the source and background.
  The quadrant-like spots at the top (in CCD3) and the bottom (in CCD4)
  are identified with the Mn K events for energy calibration.  
}\label{fig:Crabimage}
\end{figure}

The image of the Crab nebula taken with this combined mode is shown in
figure~\ref{fig:Crabimage}. 
The Crab nebula is imaged at around the optical axis on CCD2.
The brightness at the off-axis area in CCD1 and CCD2 is much lower than
that in CCD3 and CCD4 due to the short effective exposure of the burst mode. 
The diffuse structure seen in CCD3 and CCD4 is due to the tail of
the point spread
function (PSF) together with the dust scattering in the interstellar
matter. The fine structure is clearly seen owing to the efficient
exposure of ``Full Window + No Burst" mode. 

With the ``Full Window + 0.1-s Burst" mode, it is found that the core of
the Crab nebula image is still piled up by 20\%. 
Photons are not only registered during the actual integration interval
but also during the vertical transfer.
These so called Out-of-Time (OoT) events are seen like a bar along
the vertical transfer direction in figure~\ref{fig:Crabimage}. 
The OoT events are well known to experience much less pile-up. 
The magnitude of this effect scales with the ratio of readout time and integration time.
% Table~\ref{tab:burstparam} summarises the relevant parameters. 
The pile-up fraction of the OoT events is negligibly small as it is
reduced by two orders of magnitude compared with that of the integrated events. 
All the below spectral and timing analyses of the Crab data
are performed using these OoT events.

\begin{figure}
  \begin{center}
   \includegraphics[width=\linewidth]{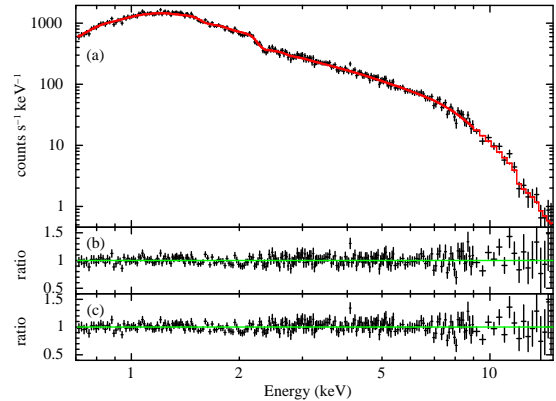}
  \end{center}
  \caption{(a) Crab nebula spectrum with the best-fit model (red solid line).
  The correction factor is applied to the effective area.
  (b) Residuals of the data from the best-fit model in (a).
  (c) Same as (b) but the correction factor is not applied to the effective area. 
}\label{fig:Crabspec}
\end{figure}

The spectrum of the Crab nebula including the pulsar contribution
is plotted in figure~\ref{fig:Crabspec}.
The event threshold was set to 100~ch that
corresponds to 0.6~keV during the Crab observation to
minimize the effect of the cross-talk issue.
Then we utilize the events above 0.7~keV for the spectral analysis
avoiding those just above the threshold.
We extract the Crab nebula spectra from the green rectangular
region (figure~\ref{fig:Crabimage}) that
covers a half of the OoT events. 
The count rate of the spectra is 2,569~counts~s$^{-1}$.
%The exposure time is as short as 21.3~s.
It takes 18.432~ms to transfer charges throughout
the extracted region in each frame. The total exposure time summed
over all frames is 21.3~s.
The binning factor is 2--95~ch per bin depending on the energy. 
Each bin contains 36--649~counts below 11.8~keV.
Five or more photons are contained above 11.8~keV.
The background spectrum is extracted from the adjoining rectangular
region in cyan (figure~\ref{fig:Crabimage}).
The count rate of the background spectrum is 57~counts~s$^{-1}$ that is about 2.2\%
of that of the source spectrum and is found to be dominated
by the X-rays from the outskirts of the PSF of the Crab nebula. 
 
\begin{figure}
 \begin{center}
    \includegraphics[width=\linewidth]{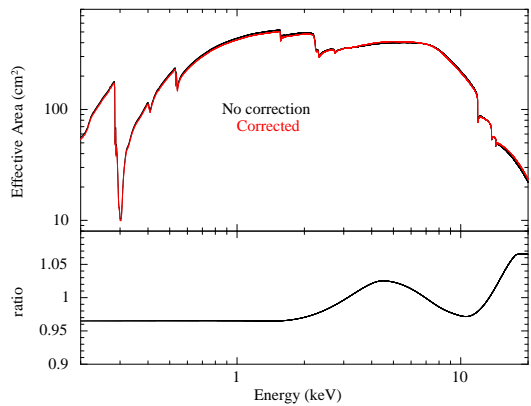}
%    \includegraphics[width=0.7\linewidth,bb=73 43 517 699]{sxi_qe_effarea_20170620.pdf}
%\vspace{15mm}
 \end{center}
   \caption{SXI effective area at on-axis. Black and Red lines correspond
   to the areas that are calculated without and with the correction factor, respectively.
   The bottom panel show their ratio. } 
\label{fig:effectivearea}
\end{figure}

We fit the spectrum with a model composed of a power-law with
photoelectric absorption (\texttt{phabs} with the absorber abundance
fixed to solar).
%We fit the spectrum with a model composed of a power-law with
%photoelectric absorption. 
The result is summarized in table~\ref{Crabspec},
and is shown in figure~\ref{fig:Crabspec}.
In the fit, we adopt two kinds of the effective area
(so-called ancillary response file : arf).
Both are calculated
using the standard ftools \texttt{aharfgen} \citep{2016SPIE.9905E..14A}
assuming a point source,
but with and without the correction (or fudge) factor
(figure~\ref{fig:effectivearea}).
We do not collapse the two-dimensional PSF to
one dimension because the PSF shape exhibits little energy dependence
(see figure~2 in \cite{2016JATIS...2d4001S}).
It is found that the effective area calculated by \texttt{aharfgen}
has some discrepancy with that measured on ground
\footnote{https:\slash\slash{}heasarc.gsfc.nasa.gov\slash{}docs\slash{}hitomi\slash{}calib\slash{}caldb\_doc\slash{}asth\_sxt\_caldb\_fudge\_v20161223.pdf}.
The correction factor is modelled as a ratio of effective areas
between the output of \texttt{aharfgen}
and that measured on ground.
The ground measurements were made at 1.5, 4.5, 8.0, 9.4, 11.0
and 17.5~keV.
A spline interpolation was applied inbetween the energies.

\begin{table*}[ht]
\caption{Best-fit parameters of the absorbed power-law model for the Crab spectra. 
%{\bf The PARAMETERS WILL BE CHANGED AND MUST BE CONFIRMED. THEY SHOULD BE CONSISTENT WITH THE OTHER PAPERS.}
}
\begin{center}
\begin{tabular}{llcccc}
\hline\hline
%\rule[-1ex]{0pt}{3.5ex}   & Correction factor & $N_{\rm H}^{\ast}$ &  Photon index $\Gamma$ & Flux$^\dagger$ & \\
\rule[-1ex]{0pt}{3.5ex}   & Correction factor & $N_{\rm H}^{\ast}$ &  Photon index $\Gamma$ & Flux$^\dagger$ & Red. $\chi^2$ (d.o.f.)\\
\hline
%$\chi ^2$ fit & & & & & Red. $\chi^2/$d.o.f. \\
%\hline
\rule[-1ex]{0pt}{3.5ex}    & Applied & 3.9~$\pm$~0.1 & 2.14~$\pm$~0.02 & 2.06~$\pm$~0.02 & 1.024 (245)  \\
\rule[-1ex]{0pt}{3.5ex}    & Not Applied & 3.8~$\pm$~0.1 & 2.11~$\pm$~0.02 & 2.06~$\pm$~0.02 & 1.072 (245)  \\
\hline
%\rule[-1ex]{0pt}{3.5ex}   c-statistics fit & & & & & C/d.o.f.\\
%\hline
%\rule[-1ex]{0pt}{3.5ex}    & Applied & 3.9~$\pm$~0.1 & 2.15~$\pm$~0.02 & 2.05~$\pm$~0.02 & 2140.09/2379 \\
%\rule[-1ex]{0pt}{3.5ex}   & Not Applied & 3.9~$\pm$~0.1 & 2.12~$\pm$~0.02 & 2.05~$\pm$~0.02 & 2159.94/2379  \\
%\hline
\end{tabular}
\end{center}
$ \ast $ Equivalent hydrogen column density in units of $10^{21}$~cm$^{-2}$. \\
%† Power-law normalization in units of photons cm$^{-2}$ s$^{-1}$ keV$^{-1}$ at 1 keV. \\
$ \dagger $ Unabsorbed energy flux in units of 10$^{-8}$~erg~cm$^{-2}$~s$^{-1}$ in the 2--10~keV band. \\
%‡ Unabsorbed energy flux in units of 10$^{-8}$ cm$^{-2}$ s$^{-1}$ in the 2--10 keV band. \\
\label{Crabspec}
\end{table*}%

%In doing the fit, we used two different statistics, $\chi^2$ \cite{Peterson1900} and
%c-statistics\cite{1979ApJ...228..939C}. 
%No binning of the spectra was made for the c-statistics fitting whereas a grouping
%is set for the $\chi^2$ fit. 
%The binning factor is 2--95~ch per bin depending on the energy. 
%The grouping is not equally spaced and is 2 from 2 to 250,  
%4 from 250 to 550,  
%7 from 550 to 900,  
%15 from 900 to 1500,  
%60 from 1500 to 2400,  
%45 from 2400 to 4000,   and
%95 from 4000 to 4095 channels,  respectively. 

The best-fit parameters of $\Gamma$ and flux in table~\ref{Crabspec}
%$\Gamma$~$\sim$~2.1 and
%flux $\sim$~2.0~$\times$~10$^{-8}$~erg~cm$^{-2}$~s$^{-1}$
are
in agreement with those of $\Gamma$~=~2.10~$\pm$~0.03 and
flux~=~(2.2~$\pm$~0.2)~$\times$~10$^{-8}$~erg~cm$^{-2}$~s$^{-1}$
by \citet{1974AJ.....79..995T}. 
This agreement supports the suggestion by \citet{2017ApJ...841...56M} 
that the X-ray variability on yearly timescales of the Crab nebula
is not part of a long term trend, but instead results from fluctuations
around a steady mean.

The effective area of the SXI is calibrated by combining the
ground-based measurements.
The measurements of the effective area of the mirror assembly,
the transmissivity of the assembly filter (i.e., thermal shield),
the transmissivity of the detector filters and the quantum efficiency
of the detector are already reported separately
\citep{Iizuka17, 2011SPIE.8147E..04T, 2014SPIE.9144E..5DK,
2012AIPC.1427..255K, 2008SPIE.7011E..0QT}.
The good agreement of the spectral parameters to the numbers in \citet{1974AJ.....79..995T}
suggests that the effective area of the soft X-ray imaging system in orbit is
consistent with those measured on ground. 

\subsection{Contamination}\label{ssec:contami}

%affects of the contaminants to CCD
Recently, contaminants on the light path of X-ray optics
are recognized to be issues in terms of the quantum efficiency in
the low energy band. Volatile materials desorbed from
the various components in the spacecraft may be adsorbed
by the cold surface such as the CCDs.
According to the extensive investigation through the ground
test of Suzaku, most problematic
contaminants is a macromolecular compound often used in plastics
\citep{Anabuki05}. Furthermore, the in-orbit
calibration of the XIS showed that the contaminant composition
evolved throughout the mission. Thus, continuous monitoring
of the contamination thickness is essential to
properly calibrate quantum efficiency. 
%Especially the outgassing from the structure inside the satellite
%is generally severe in the initial operation.
Monitoring the contamination on the SXI also gives criterion to open the
gate valve of the SXS in the initial operation.

%about RX J1856
Hitomi observed RX~J1856.5--3754
twice, a week apart, to monitor any possible accumulation
of such a contaminant. Although Hitomi aimed at the target
direction precisely for both pointings, the signal events were
detected only for the SXI because the quantum efficiency of the SXS
below 2~keV is too low to detect the emission from the target
with the gate valve closed. The emission from the target
is too soft for the hard X-ray imaging system
\citep{2014ApOpt..53.7664A, 2016NIMPA.831..235S}.

The X-ray spectrum of RX~J1856.5--3754 is well reproduced with
the two-temperature blackbody emission model \citep{2006A&A...458..541B}.
%The temperatures measured by an observer at infinity
%are $kT_{h}^{\infty}$~=~62.83~$\pm$~0.41~eV and $kT_{l}^{\infty}$~=~32.26~$\pm$~0.72~eV
%with each radius of $R_{h}/d$~=~0.0378~$\pm$~0.0003~km~pc$^{-1}$
%and $R_{l}/d$~=~0.1371~$\pm$~0.0010~km~pc$^{-1}$, respectively
%\citep{2006A&A...458..541B}.
Because the spectrum has been
stable over years \citep{2012A&A...541A..66S}
with small fraction of the pulsation (1.2\%)
in the 0.15--1.2~keV band \citep{2007ApJ...657L.101T},
RX~J1856.5--3754 has often been used to
calibrate other soft X-ray instruments.
Although the excess of the emission over the conventional
two-temperature blackbody model around 1~keV
is recently reported \citep{2017PASJ...69...50Y},
we do not introduce the excess because of the
limited photon statistics.

\begin{figure}
  \begin{center}
  \includegraphics[width=\linewidth]{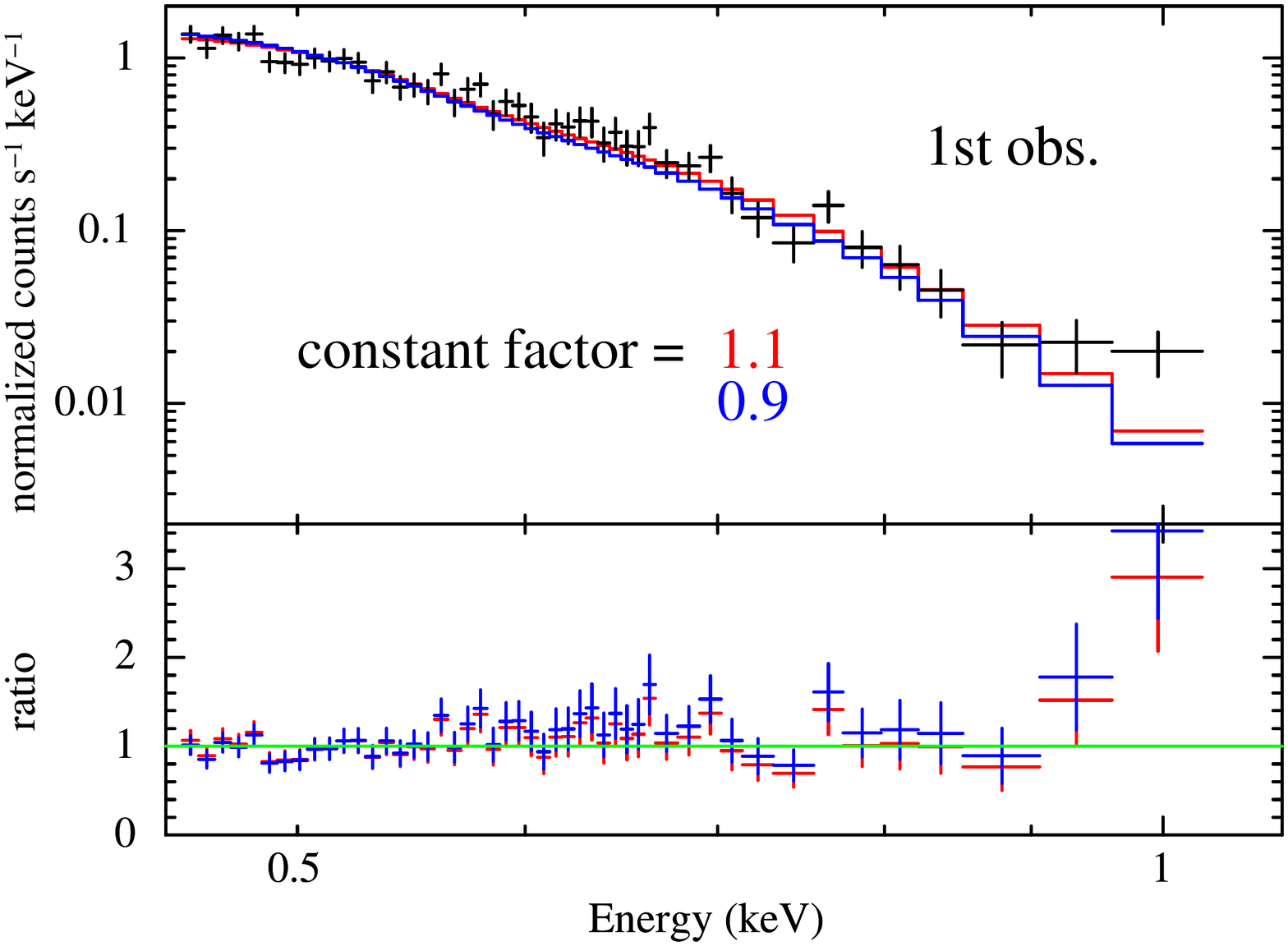}\\
  \includegraphics[width=\linewidth]{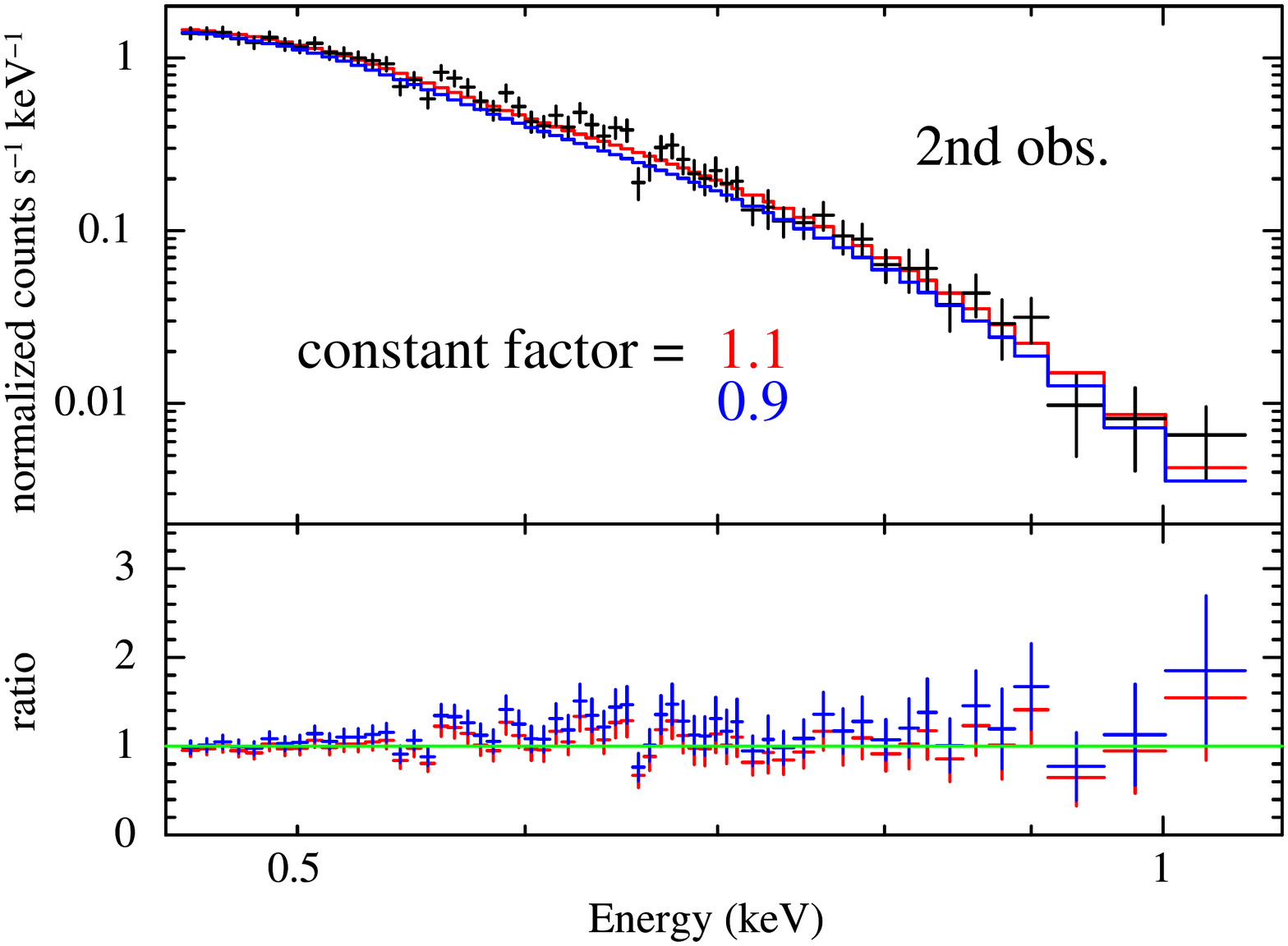}
  \end{center}
 \caption{(Top) RX~J1856.5--3754 spectrum at the first observation with background
 subtracted. Red and blue lines correspond to the best-fit model with a constant
 factor of 1.1 and 0.9, respectively. 
 (Bottom) Same as the top panel, but for the second observation.
 }\label{fig:contamination}
\end{figure}

Figure~\ref{fig:contamination} shows the background-subtracted
spectra from the two observations.
We extract the source spectra from a circular region with a radius of \timeform{90"}
and the background spectra is chosen from the entire region of the on-axis segment
excluding a circular source region with a radius of \timeform{5'}.
The difference of the effective area between the source and background regions
is corrected as well as the difference of the area between the regions.
%Although the event threshold is set to 40~ch in these observation,
%the events that spread over four pixels have the minimum PH of 75~ch
%because we set the split threshold to 15~ch. Then we set the lower edge of
%the energy in the spectral fit to be 0.45~keV that corresponds to the pulse
%height of 75~ch.

We fit the spectra with the model of
\texttt{phabs*\{bbodyrad+bbodyrad\}*constant}, where
the \texttt{constant} factor is introduced to absorb the calibration
uncertainty of the effective area in the soft energy band.
To estimate systematic errors,
we perform the spectral fit by fixing the constant factor
to 0.9 and to 1.1
that are conventionally used as a uncertainty range of the absolute
effective area
and are exemplified in section~\ref{ssec:QEandEA}.
The model equivalent hydrogen column density,
%($N_{\rm H}$~=~1.1~$\times$~10$^{20}$~cm$^{-2}$),
blackbody temperatures, and radii for each temperature component
are adopted from \citet{2006A&A...458..541B}.
Additional multiplicative model of the column densities due to the contaminants
is applied.
Possible compositions of the contaminants are neutral H, C, and O.
However, we allow only the column density of C ($N_{\rm C}$) to vary
from the statistical point of view.

Table~\ref{tab:rxj1856} summarizes
the best-fit parameters from the two observations.
Although $N_{\rm C}$ is consistent with zero or negative in the case of the
constant factor of 0.9, it is significantly detected
when we set the factor to 1.1. Considering these facts,
we conclude that $N_{\rm C}$ is less than
9.4~$\times$~10$^{17}$~cm$^{-2}$ and 5.5~$\times$~10$^{17}$~cm$^{-2}$
for the first and second observations, respectively, which corresponds
to the mass column density of C ($\rho_{\rm C}$~=~12~$\times$~$N_{\rm C}$/$N_{\rm A}$)
of 18.8~$\mu$g~cm$^{-2}$ and 10.8~$\mu$g~cm$^{-2}$.
The pre-launch requirement of the SXI for
the limit of any contaminant is
$\rho_{\rm C}$~$<$~10~$\mu$g~cm$^{-2}$. 
Follow-up calibrations would have made the results more precise.
%The thickness of contaminant is 0.11~$\mu$m assuming a specific gravity of unity.
Another important fact is that $N_{\rm C}$ did not show significant increase
during the interval between the two observations, which assures the opening
safety of the SXS gate valve.

\begin{table}[ht]
\caption{Best-fit parameters for RX~J1856.5--3754 spectra.}
\label{tab:rxj1856}
\begin{center}       
%\begin{threeparttable} % 2017.4.11
\begin{tabular}{lcc} %% this creates two columns
%% |l|l| to left justify each column entry
%% |c|c| to center each column entry
%% use of \rule[]{}{} below opens up each row
\hline\hline
Parameters  & \multicolumn{2}{c}{First Observation} \\
%\rule[-1ex]{0pt}{3.5ex}  $N_{\rm H}$ (cm$^{-2}$) & \multicolumn{4}{c}{1.1$\times$ 10$^{20}$}  \\
%\hline
%\rule[-1ex]{0pt}{3.5ex}  kT$_{h}^{\infty} {\rm(eV)}$ & \multicolumn{4}{c}{62.83}   \\
%\hline
%\rule[-1ex]{0pt}{3.5ex}  $R_{h}^{2}/D_{10}^{2}$\tnote{*2} & \multicolumn{4}{c}{1.429$\times$10$^{5}$} \\
%\hline
%\rule[-1ex]{0pt}{3.5ex}  kT$_{l}^{\infty} {\rm(eV)}$ & \multicolumn{4}{c}{32.26}   \\
%\hline
%\rule[-1ex]{0pt}{3.5ex}  $R_{l}^{2}/D_{10}^{2}$ & \multicolumn{4}{c}{1.880$\times$10$^{6}$}   \\
\hline
Constant factor$^\ast$ (fixed) & 0.9 & 1.1   \\
$N_{\rm C}$ (10$^{17}$~cm$^{-2}$) & 0.8$^{+1.5}_{-1.4}$ & 7.9$^{+1.5}_{-1.4}$   \\
Reduced $\chi^{2} ({\rm d.o.f.})$ & 1.305 (47) & 1.032 (47)   \\
\hline
 & \multicolumn{2}{c}{Second Observation}  \\
\hline
Constant factor$^\ast$ (fixed) & 0.9 & 1.1   \\
$N_{\rm C}$ (10$^{17}$~cm$^{-2}$) & --2.6$^{+1.0}_{-0.9}$ & 4.5$^{+1.0}_{-1.0}$   \\
Reduced $\chi^{2} ({\rm d.o.f.})$ & 1.340 (59) & 0.939 (59)   \\
\hline
\end{tabular}
%\begin{tablenotes}\footnotesize
%\multicolumn{4}{l}{[*1] Parameters without errors are fixed to the values in \citet{2006A&A...458..541B}.}\\
%\multicolumn{4}{l}{[*2] $R$ is the source radius in km and $D_{10}$ is the distance to
%the source in units of 10~kpc.}\\
%\multicolumn{4}{l}{[*3] Constant factor is introduced to take charge of the calibration
%uncertainty of the effective area in the soft energy band.}\\
%\end{tablenotes}
%\end{threeparttable}
\end{center}
$\ast$ Constant factor is introduced to take charge of the calibration
uncertainty of the effective area in the soft energy band.  \\
\end{table} 

\subsection{Point Spread Function}

PSF of a focusing telescope system is usually
calibrated using a bright and point-like source. 
The brightness reduces the uncertainty of the background subtraction
at its tail whereas the point-like emission works to resolve its sharp core.   
Due to the limited numbers of the observed targets of Hitomi,
any bright and point-like source was not observed. 
Instead, we use an on-pulse image (the pulsar image) of the Crab. 
%The in-flight calibration of the PSF using the Crab pulsar is presented below. 

\subsubsection{Extraction of the On-pulse Component}

OoT events give a huge merit in terms of the timing resolution. 
%The OoT events are recorded when the pixel just passes the focus. 
The row position (DETX) can be used as a time tag of events
with the time unit of 57.6~$\mu$s that corresponds to
the readout clock per row. 
However, the focus of the SXT-I is blurred with its PSF.
The tagged time is then smoothed by the PSF (\timeform{0.7'} or 1.14~mm in
half power widths (HPWs), see below).
The effective time-resolution becomes $\sim$~1.3~ms.
%($=$ 1.14~mm$/$50[$\mu$m$/$row] $\times 57.6$[$\mu$s/row]) . 

\begin{figure}
 \begin{center}
 \hspace{-25mm}
  \includegraphics[width=0.7\linewidth, bb=0 0 500 700]{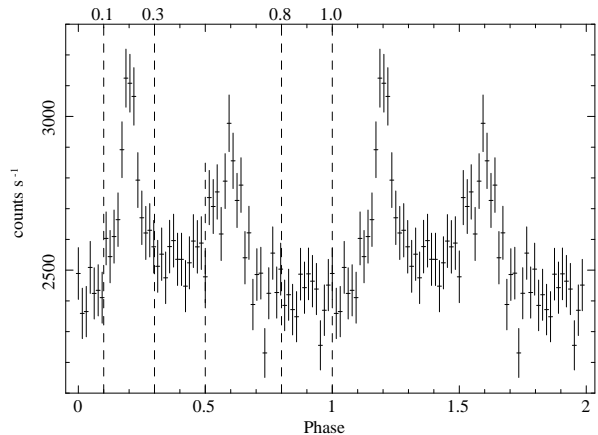}
  \end{center}
  \vspace{2mm}
  \caption{Folded light curve of the OoT events of the Crab nebula.
  The phase zero is taken arbitrarily.
}\label{fig:Crabfoldcurve}
\end{figure}

Figure~\ref{fig:Crabfoldcurve} shows a folded light curve of
the Crab in the total band.
We adopt a period of 33.7204626~ms at the epoch
of 17472~d (TJD) that is obtained with the SXS \citep{maeda17}. 
The timing analysis of all the Hitomi instruments is
comprehensively summarized in other papers (\cite{2016SPIE.9905E..3UL};
Koyama et al. in preparation).
The pulse is clearly detected in the folded light curve.

The pulses have widths of 0.1--0.2 phase that corresponds to 3--6~ms.
The emission region of the pulsed component must be smaller than
3--6~$\times$~$10^{-3}$ times the speed of light ($\approx$~1--2~$\times$~10$^{7}$~cm).
If we assume that the distance to the Crab nebula is 2~kpc, the source
of the pulsed component (i.e., the Crab pulsar) 
is securely regarded as a point-like source.

Dust scattering is known to make a halo structure around the source
\citep{2003ARA&A..41..241D}.
Once the emission from the pulsar is scattered, it experiences a time lag
due to a detour of the arrival pass and makes the emission unpulsed.
Therefore, the pulsed component is an ideal point-like source that is
free from the dust scattering halo. 
Owing to its large brightness of the Crab pulsar, the pulsed component
is used as an in-flight calibration of the PSF.  

\subsubsection{One Dimensional Point Spread Function}

\begin{figure}
\vspace*{20mm}
 \begin{center}
 \hspace{5mm}
 \includegraphics[width=\linewidth, bb=0 150 720 340]{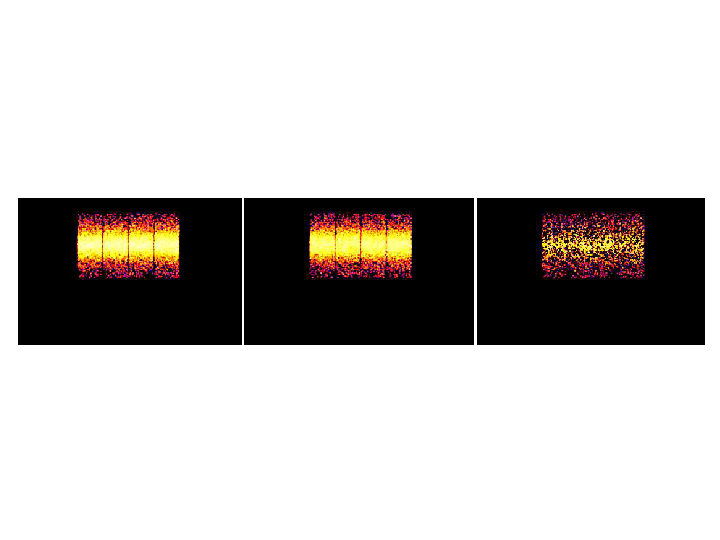}\\
%    \includegraphics[width=0.6\linewidth, angle=-90, bb=62 40 511 717]{comp_arcmin_pulsar_nknw_on-off_pro_dety_0416_bin4_2_plt.pdf}\\
%\vspace*{2mm}
\hspace*{10mm}On pulse\hspace*{9mm}Off pulse\hspace*{5mm}On-minus-off pulse\\
    \includegraphics[width=\linewidth]{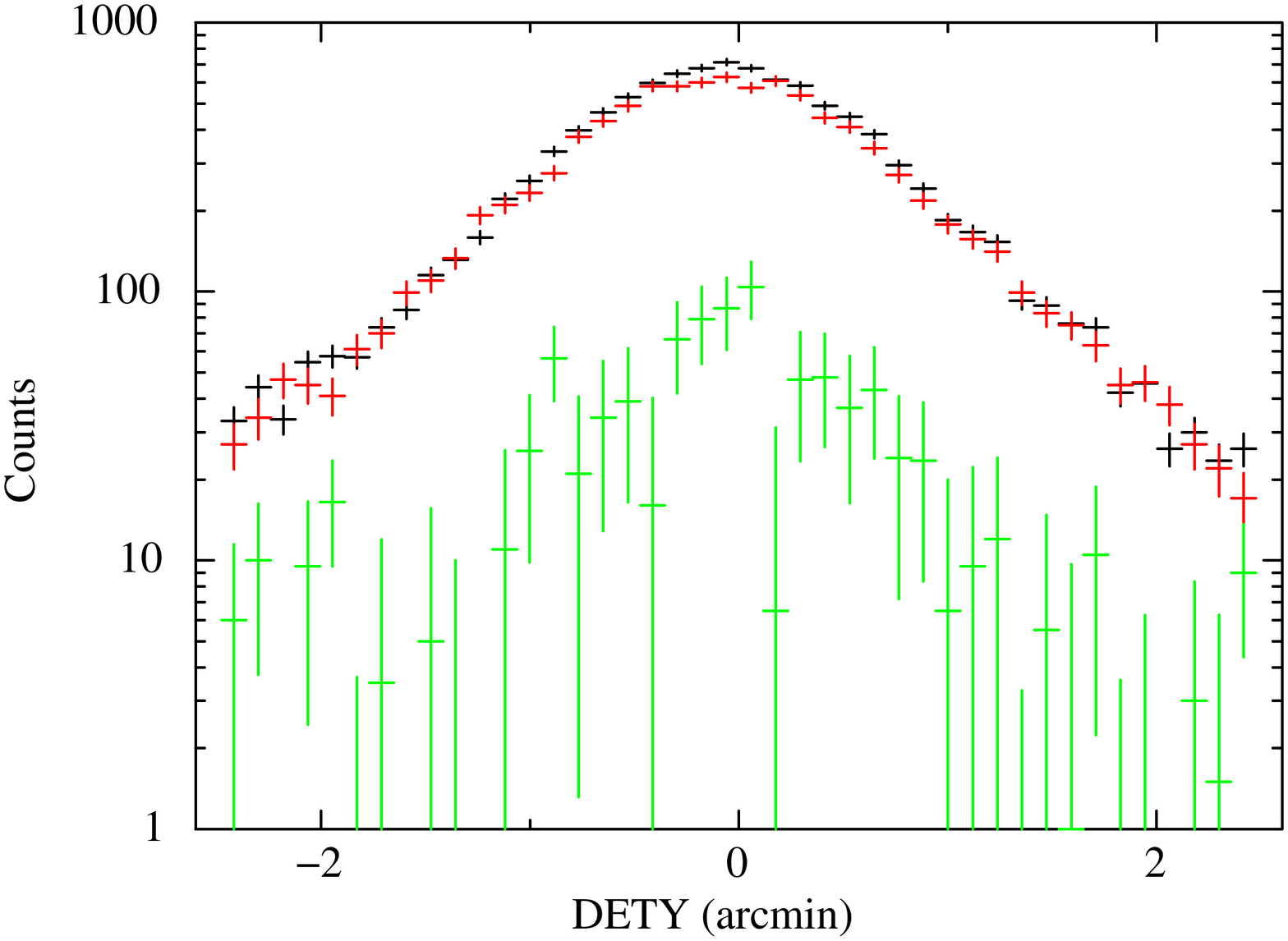}\\
  \end{center}
%\vspace*{10mm}
  \caption{(Top) Out-of-Time images of the Crab nebula observation in the DET
  coordinate for the on-pulse
  (left), the off-pulse (middle) and the on-minus-off pulse (right).
  (Bottom) Projection along the DETX direction.
  The black, red and green data correspond to the projections using the pulse phase of on,
  off and on-minus-off, respectively.}
  \label{fig:OoTimage}
\end{figure}

Figure~\ref{fig:OoTimage} shows the 0.6--15~keV band images of the on-pulse,
the off-pulse and the on-minus-off pulse. 
The images are extracted from the green region
shown in figure~\ref{fig:Crabimage}.
In figure~\ref{fig:Crabfoldcurve}, we
define the phase intervals 0.1--0.3 and 0.5--0.7 as the on-pulse phase.
%identify the on-pulse phase,
%the 1st in 0.1--0.3 phase and the 2nd in 0.5--0.7 phase. 
The phase interval 0.8--1.0 is adopted as an off-pulse phase.  
The effective exposure times of the on- and off-pulse images
are 8.6~s and 4.3~s, respectively.
The on-minus-off pulse image is made by subtracting the off-pulse
image from the on-pulse.
The on-minus-off pulse image appears narrower in the DETY direction.  
It is because the on-minus-off pulse image originates from the pulsar radiation,
i.e., the point-like emission. 

\begin{figure}
 \begin{center}
 \includegraphics[width=\linewidth]{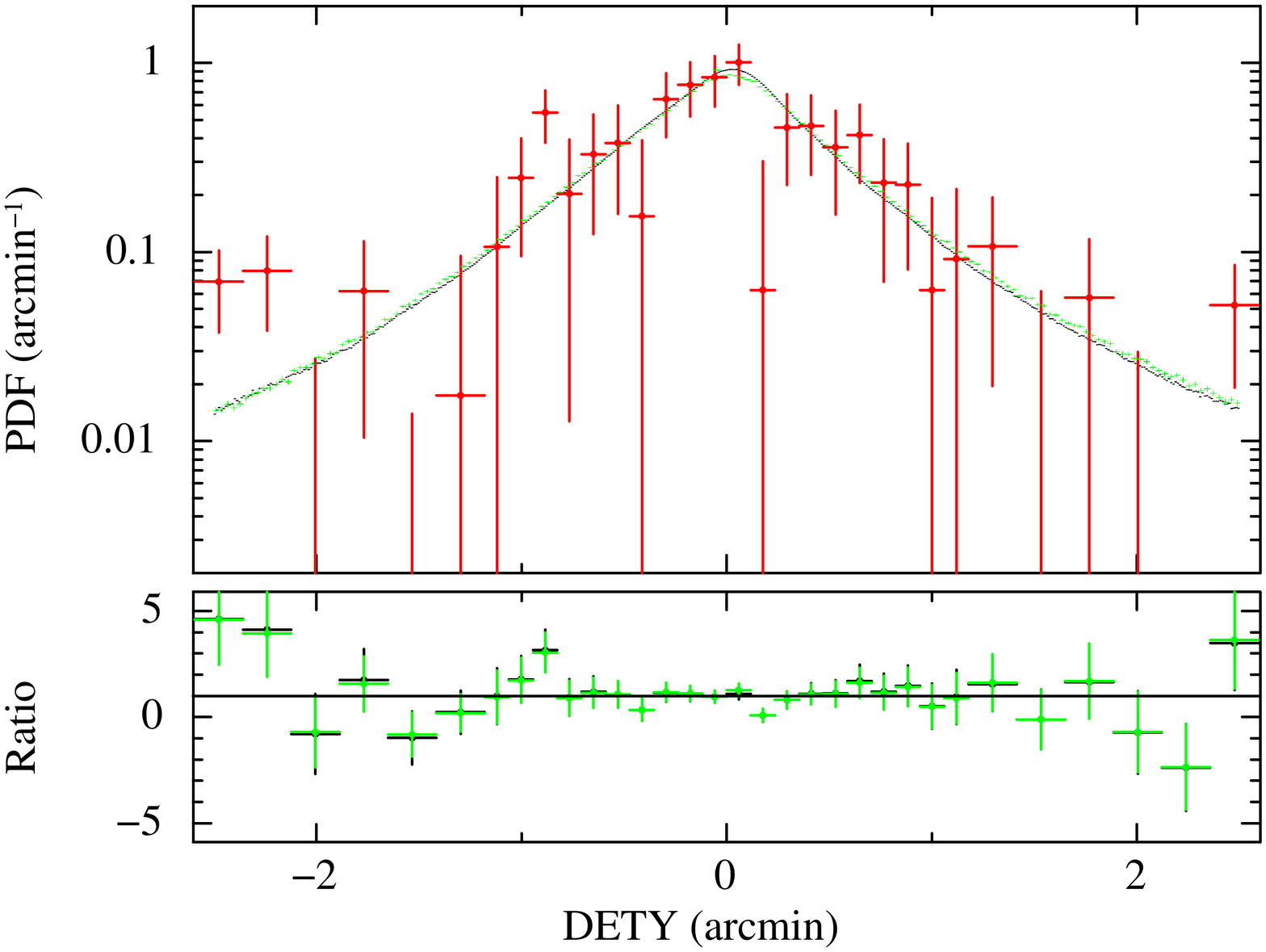}\\
% \includegraphics[width=0.6\linewidth, bb=62 40 511 717]{comp_kuramodel_pro_ratio_plt.pdf}\\
%\vspace*{2cm}
 \includegraphics[width=\linewidth]{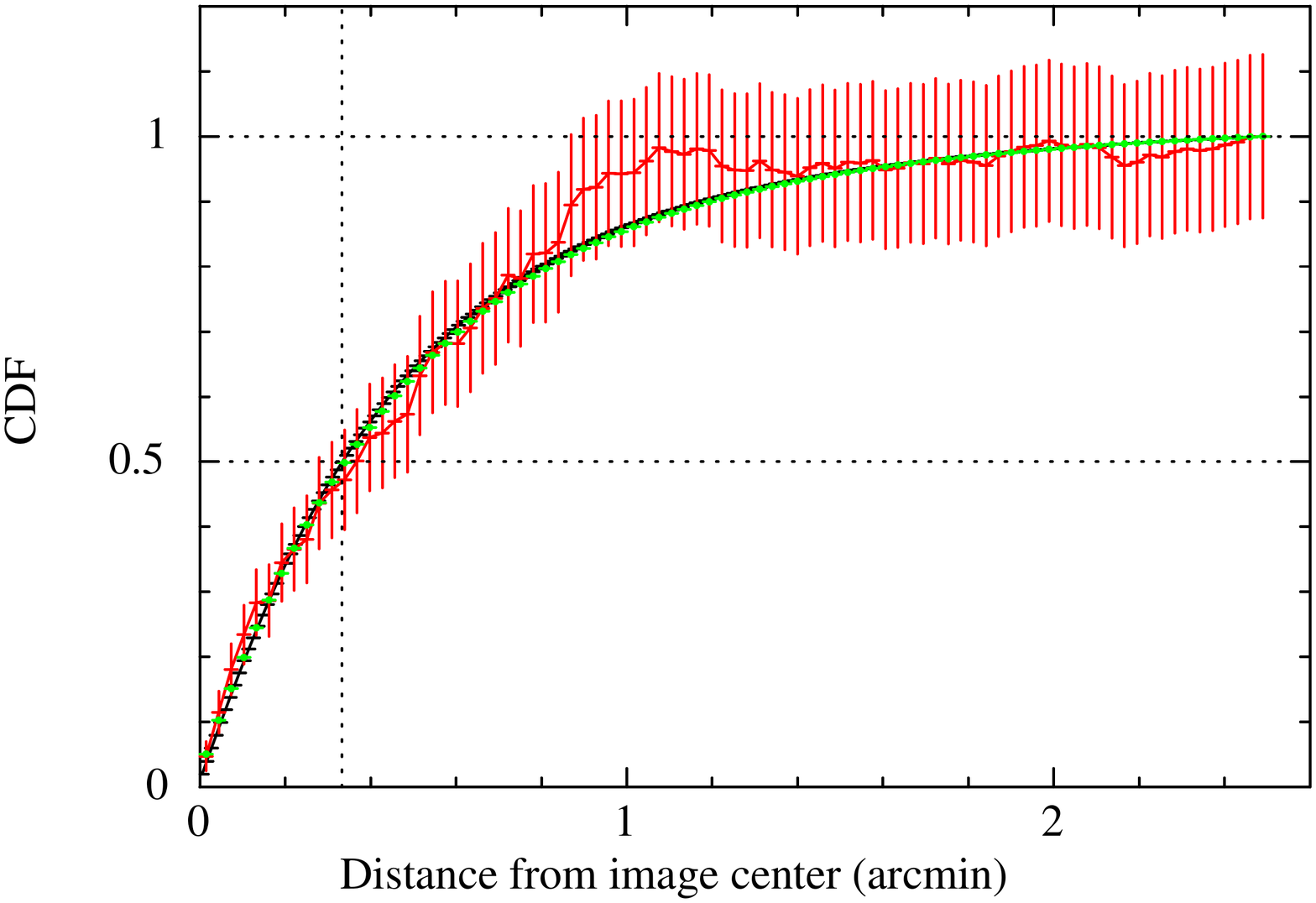}\\
% \includegraphics[width=0.6\linewidth, bb=62 40 511 717]{comp_hpw_0606_plt.pdf}\\
%\vspace*{1cm}
  \end{center}
  \caption{(Top) One-dimensional PSF (or PDF: probability density function)
  of the Crab nebula's on-minus-off pulse image (red).
  The ground-based measurements (black) and the ray-tracing output (green)
  are overlaid. Ratios with respect to the observed data are also shown.
  (Bottom) Cumulative distribution function (CDF) from the peak of the
  one-dimensional PSF with the same color coding as the top panel.
  The vertical dash line follows the half width of the half power,
  i.e., the half of the HPW.}
  \label{fig:1DPSF}
\end{figure}

Since the OoT trail along the direction of DETX remained,
we make a histogram of the events as a function of DETY.
This histogram corresponds to a one-dimensional PSF
and is shown in figure~\ref{fig:1DPSF}.
%A cumulative distribution function from the peak of the one-dimensional
%PSF is also plotted in figure~\ref{fig:1DPSF}.
To compare the ground measurements with the in-flight PSF,
we simply collapse the image taken on ground down to one
dimension along the direction of DETX and make a one-dimensional
PSF for the ground measurements. The cumulative distribution function
from the peak of the one-dimensional PSF is plotted in figure~\ref{fig:1DPSF}
as well as the output of ray-tracing.
HPWs are calculated using the cumulative distribution
function and are listed in table~\ref{1DPSF}.
Within the limited statistics of the data, the HPW of the Crab pulsar
is consistent with that of the ground measurement at 4.51~keV and that
of the ray-tracing used in the calculation of the response function of the SXT-I.
It can be concluded that the in-flight PSF  of the SXT-I is consistent
with the on-ground measurements. This indicates that the HPD of the SXT-I is
$\sim$~\timeform{1.3'} in orbit that is the number we measured on ground.

\begin{table}[h]
\caption{Half power widths calculated from the one dimensional PSFs.}
\begin{center}
\begin{tabular}{lccc}
\hline
\hline
Data 			       & HPW (arcmin)  & References$^\ddagger$ \\
\hline
The Crab	pulsar$^\ast$	&  $0.74\pm0.12$     & This work. \\ 
Ground measurements$^\dagger$	&  $0.67\pm0.01$     &  (1), (2) \\ 
Ray-tracing	&  $0.68\pm0.02$     &  (3)  \\
\hline
\end{tabular}
\end{center}
\label{1DPSF}
$\ast$ The value is derived from the on-minus-off pulse image (see the text for details).  \\
$\dagger$ Monochromatic energy of 4.51 keV. \\
$\ddagger$ (1)\citet{2016JATIS...2d4001S}; (2)\citet{Iizuka17}; (3)\citet{Yaqoob2017a}.\\
%$\ddagger$ (1)\citet{2016JATIS...2d4001S}; (2)\citet{Iizuka2017a}; (3)\citet{Yaqoob2017a}\\
\end{table}%

\subsection{Non X-ray Background}

Low and stable background is essential for precise observations
especially of extended targets. Hitomi adopts almost the same LEO
as Suzaku that realized low and reproducible NXB \citep{2008PASJ...60S..11T}.
Design of the camera body is another key to suppress the NXB.
Considering a result of a Monte-Carlo simulation \citep{2008SPIE.7011E..3XA},
we designed the sensor body so that the thickness of the housing is
$\geq$10~g~cm$^{-2}$.

%NXB spectrum
During the three weeks of the observations, we accumulated the events
in the period when the spacecraft pointed to the night earth and regard
those events as NXB. 
To securely extract only the NXB events, we define a screening criterion
from extended house keeping data as

\texttt{
%\begin{equation}
(SAA\_SXI\ ==\ 0)\ \&\&\ (T\_SAA\_SXI\ >\ 277)\\
\hspace{0.2\linewidth} \&\&\ (ELV\ <\ -5)\ \&\&\ (DYE\_ELV\ >\ 100).
%\end{equation}
}

\begin{figure}
 \begin{center}
  \includegraphics[width=\linewidth]{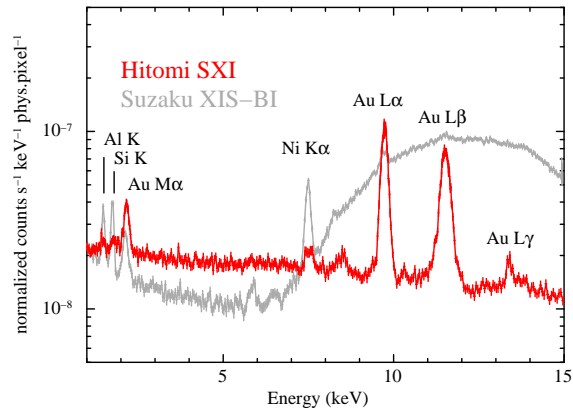}\\
 \end{center}
 \caption{SXI NXB spectrum normalized by physical pixel area.
 Events from the entire FoV are collected with the exception of
 the regions irradiated with the calibration source events and their OoT events.
 Data of Suzaku XIS-BI that have the same physical pixel area is
 also shown for comparison. The data of the XIS-BI
 is adopted from \citet{2008PASJ...60S..11T}. 
 }\label{fig:NXB_area}
\end{figure}

%NXB spectrum
As a result, the total exposure is 118.3~ks
for the on-axis segment including multiple pointings for the check of the AOCS
and excluding the Crab observation that is not
performed with the normal clocking mode of ``Full Window + No Burst".
The exposure times for the other segments are different among pixels
depending on whether the pixel was affected by the cross talk issue.
Figure~\ref{fig:NXB_area} shows the integrated NXB spectrum
normalized with respect to the area of a physical pixel.
All the good grade events are summed over the whole FoV with
the exception of the regions irradiated with the calibration sources
and those affected by OoT Mn K events. The region
with 360 $\le$ DETY $\le$ 1450 is regarded to be free from
the OoT events. 

%detail of the spectrum
Emission lines from Au L$\alpha$, Au L$\beta$,
Au M$\alpha$ and weak lines from Al K,
Si K, Ni K$\alpha$, and Au L$\gamma$ are seen.
Primarily the emission lines from Au originates
from the inner surface of the housing. The surface of the CCD
package and the cold
plate are also vapor deposited by Au. Ni is under the Au layer
of the housing. The OBL on the surface of the CCD
and the CBF consist of Al
and hence they are responsible for the Al emission line.
The Si line is emitted from the inside of the CCD wafer.
The NXB spectrum of XIS-BI has a hump above 6~keV, 
which limits the sensitivity
in the energy band. In contrast, the NXB spectrum of the SXI
shows no hump in the energy band up to 15~keV. 
The Ni K$\alpha$ line at 7.5~keV is weaker than
that of the XIS because of a thicker Au layer vapor deposited
onto the inner surface of the housing.
This allows us to obtain a better estimation
of the spectral parameter from celestial objects
such as supernova remnants because the centroid of the
Ni K$\alpha$ overlaps with those of Fe He$\beta$ and Ni He$\alpha$.
%while the fluxes of Au L lines are stronger than those of the XIS.

\begin{figure}
  \begin{center}
  \includegraphics[width=\linewidth]{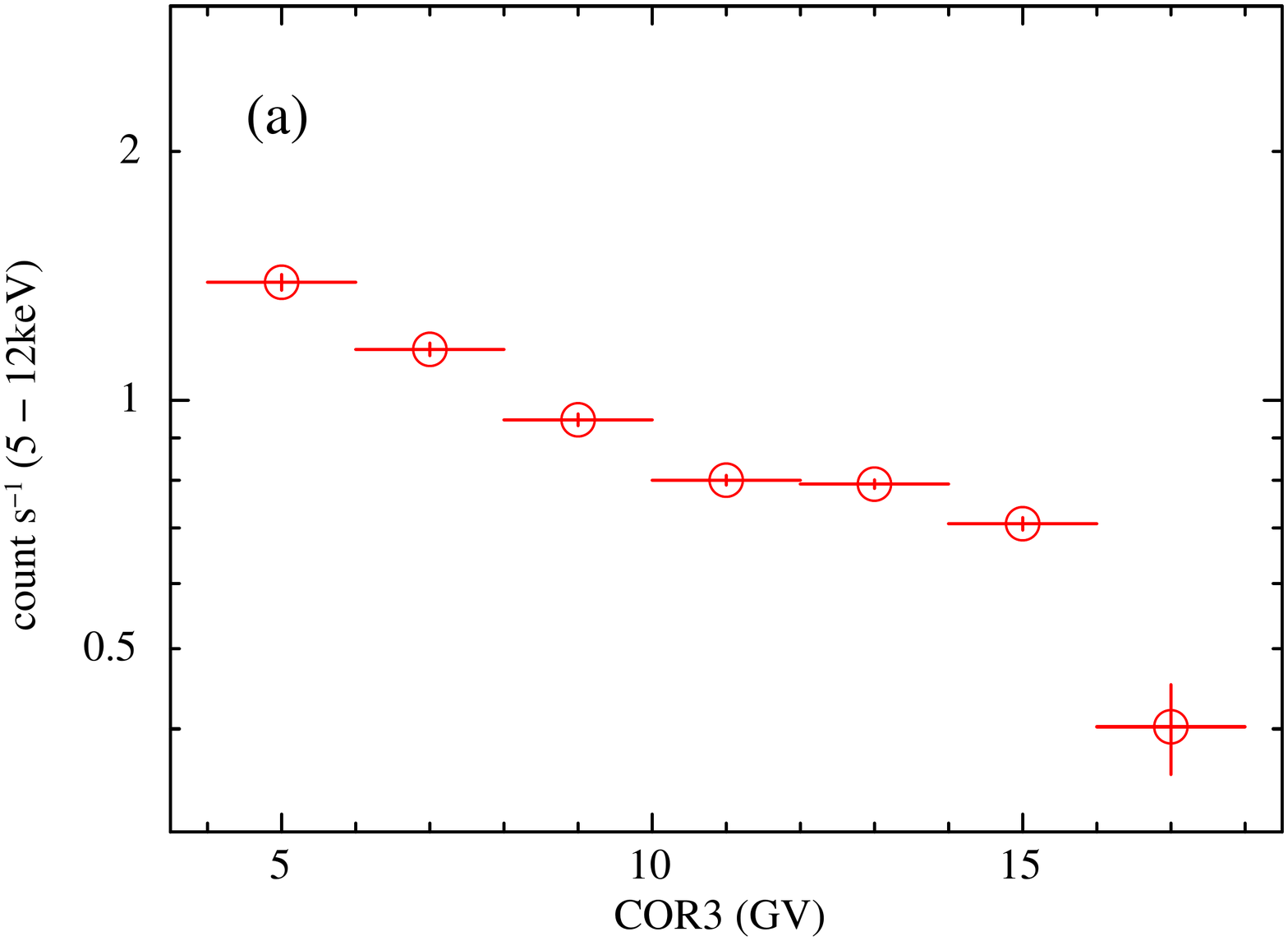}\\
  \includegraphics[width=\linewidth]{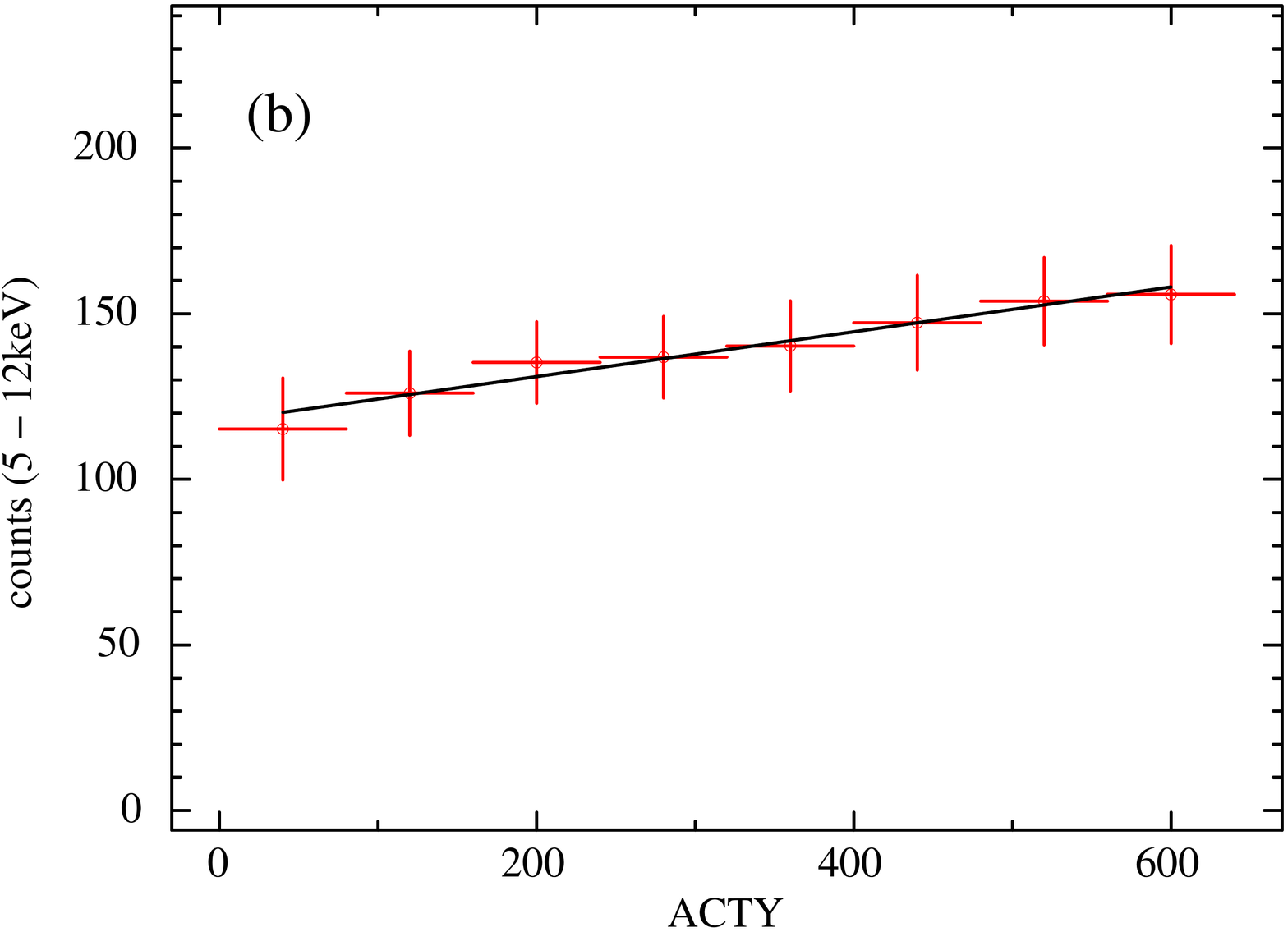}
  \end{center}
 \caption{(a) Count rate of the NXB in the energy band from 5 to 12 ~keV
 as a function of COR. (b) NXB count in the same energy band as (a)
 as a function of ACTY. Data of all the segments are summed for both
 panels.
 }\label{fig:NXB_asafuncofCORACTY}
\end{figure}

%COR and ACTY dependence
The NXB intensity in general depends on the cut-off rigidity (COR)
as investigated for the XIS \citep{2008PASJ...60S..11T}.
Figure~\ref{fig:NXB_asafuncofCORACTY}(a) shows a count rate of
the NXB
in the energy band of 5--12~keV as a function of COR that is
based on the international geomagnetic reference field model
version~12 \citep{Thebault2015}.
The NXB also depends on the position in the CCD as shown in
figure~\ref{fig:NXB_asafuncofCORACTY}(b). If the spatial distributions of
the charged particles or hard X-rays are uniform
throughout the sensor including the frame store region,
the intensity of
the NXB depends on the exposed duration at each position.
Taking into account the difference of the pixel size between
the imaging area and the frame store region, 
the intensity of the top region should be 1.7 times larger than that
of the bottom region. The measured ratio of $\sim$~1.3 is significantly lower
than that expected, which may be due to the flux difference between the imaging
area and the frame store region, or the difference of the grade branching
ratio between the two regions. 
%sxinxbgen
To derive a proper background spectrum to be subtracted from a source spectrum,
a software \texttt{sxinxbgen} takes into account
these positional and COR dependences.

%\begin{equation}
%  S_{W} = \frac{\Sigma_{i=1}^{n} T_{i} \Sigma_{j=1}^{n} A_{j} S_{i,j}}{\Sigma_{i=1}^{n}%\Sigma_{j=1}^{n}\ T_{i}\ A_{j}}
%\end{equation}

\begin{figure*}
 \begin{center}
  \includegraphics[width=\linewidth]{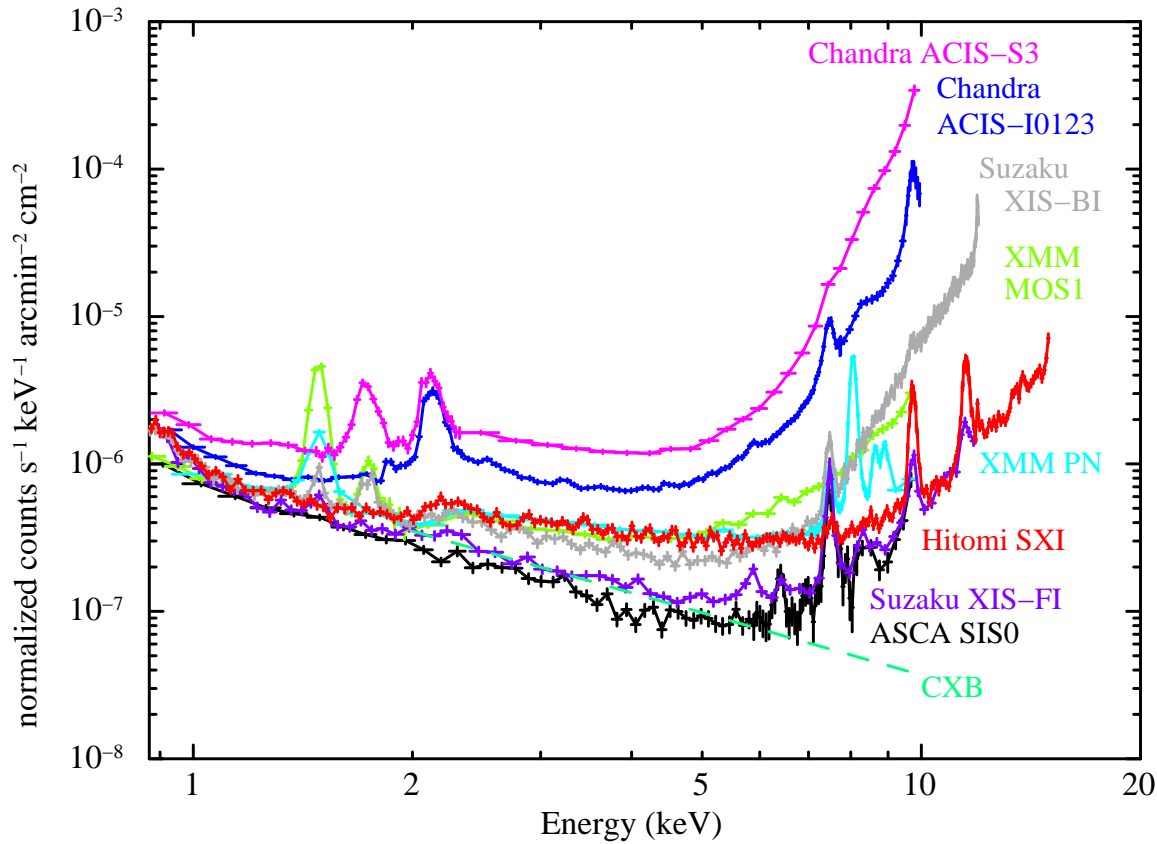}
 \end{center}
 \caption{SXI sky background spectrum normalized by effective area and solid angle
 of the FoV. CXB level and data of other instruments using X-ray CCDs are
 also shown for comparison.
 Datasets of ASCA, Chandra, and XMM-Newton are adopted from \citet{2004A&A...414..767K}.
 Suzaku XIS data are adopted from \citet{2008PASJ...60S..11T}.
 }\label{fig:NXB_4diffuse}
\end{figure*}

%diffuse sensitivity
In figure~\ref{fig:NXB_4diffuse} we compare sky background spectra among different
X-ray CCD detectors: ASCA SIS0, Chandra ACIS-\{I, S3\}, XMM-Newton \{EPIC-pn, EPIC-MOS\},
Suzaku XIS-\{FI, BI\}, and Hitomi SXI. The SXI background is extracted from
the entire FoV during the RX~J1856.5--3754 observation
with exceptions of a circular region
around the target with a radius of \timeform{150"}, calibration
sources regions, and their OoT events region.
%Other spectra are adopted from
%\citet{2004A&A...414..767K} and \citet{2008PASJ...60S..11T}.
All the spectra are normalized by the effective area of the instrument
(corresponding telescope + CCD) at its focus and by the
solid angle of the FoV. Hence it illustrates a surface brightness
of the sky background and provides a measure of sensitivity for extended
targets. The CXB intensity is set to be 
9.3~$\times$~10$^{-7}$~(E/1~keV)$^{-1.4}$~counts~s$^{-1}$~keV$^{-1}$~arcmin$^{-2}$~cm$^{-2}$
as a reference.
%Thanks to a large effective area of the SXT-I and high quantum
%efficiency of the SXI, the soft X-ray imaging system aboard Hitomi achieve better
%sensitivity than Chandra, XMM-Newton, and Suzaku/XIS-BI sensors above 6~keV.
Thanks to the high quantum efficiency of the SXI, the soft X-ray imaging system
aboard Hitomi achieve better sensitivity than Suzaku XIS-BI above 6~keV.
The sky background intensity is
5.6~$\times$~10$^{-6}$~counts~s$^{-1}$~arcmin$^{-2}$~cm$^{-2}$ in the energy
band of 5--12~keV, which is seven times lower than that of the XIS-BI.

\section{Summary}
\label{sec:summary}

In spite of the short lifetime of the mission, the soft X-ray imaging system
confirmed its function and performance.
The observation of the Perseus cluster of galaxies
demonstrated the imaging capability
over the large FoV of \timeform{38'}~$\times$~\timeform{38'}.
We measured the HPW using the on-minus-off pulse image for the OoT
events of the Crab pulsar. It is found that the HPW is consistent with
that of the ground measurement and that of the ray-tracing.
We thus conclude that the angular resolution is as good as
$\sim$~\timeform{1.3'} in orbit.
The best-fit parameters of the absorbed power-law model for the Crab spectra
is consistent with those in the literature, suggesting the correctness of
the effective area calibration.
The effective exposure time is affected by the light leak
issue primarily when the minus Z axis of the spacecraft points to the day earth.
The cross talk issue affects a part
of effective area. We would not see the issue any longer
if we had optimized the threshold for the dark level update before the loss of
the spacecraft.
After the screening of the data affected by
these two issues, the spectra of the onboard calibration source events prove
that the energy resolution at 5.89~keV is 179~$\pm$~3~eV in FWHM.
The sky background level is seven times lower than that of Suzaku XIS-BI
in the energy band of 5--12~keV.

\begin{ack}
We would like to express our appreciation for the tremendous efforts by
Yu Shiodome, Kentaro Someya, Takuro Sato, Ko Ichihara, Kazuki Tomikawa,
Naomichi Kikuchi (ISAS/JAXA), Ikuya Sakurai, Yuhki Kurebayashi (Nagoya University),
Sari Minami (Nara Women's University), Takanori Izumiya, Takashi Awaya,
Kota Okada (Chuo University), Naohisa Anabuki, Shuhei Katada (Osaka University),
Ryosaku Washino (Kyoto University), Eri Isoda (University of Miyazaki)
Tahir Yaqoob (University of Maryland Baltimore County), and
Lorella Angelini (NASA/GSFC)
in the development and calibration of the SXT-I and the SXI.
S.~Inoue, K.~K.~Nobukawa and T.~Sato are supported by the Research Fellow of JSPS for Young Scientists.
This work is supported by the JSPS KAKENHI Grant Number
JP14079204, JP15H02070, JP15H02090, JP15H03641, JP15J01845, JP15K17610, JP16H00949,
JP16H03983, JP16K13787, JP14079204, JP17K14289, JP18740110, JP20365505, JP20549005,
JP21659292, JP23000004, JP23340071, JP23540280, JP23740199, JP24684010, JP24740123,
JP25105516, JP25109004, JP25870181, JP26109506, JP26670560, JP26800102, and JP26800144.
T. Dotani acknowledges support from the Grant-in-Aid for Scientific Research on Innovative Areas
``Nuclear Matter in Neutron Stars Investigated by Experiments and Astronomical Observations''.
\end{ack}

%%%%% References %%%%%
%\bibliography{report}   % bibliography data in report.bib
\bibliography{mybibfile}
\bibliographystyle{pasjmybst}   % makes bibtex use spiejour.bst

\end{document}